\documentclass[%
 aip,
rsi,%
 amsmath,amssymb,
 reprint,%
]{revtex4-1}

\usepackage{graphicx}
\usepackage{dcolumn}
\usepackage{bm}
\usepackage{diagbox}
\usepackage{mathtools}
\usepackage{multirow}
\usepackage{nicefrac}
\usepackage{standalone}
\usepackage{float}
\usepackage[caption=false]{subfig}

\usepackage{tikz}
\usetikzlibrary{calc}
\usetikzlibrary{decorations.pathreplacing,decorations.pathmorphing,decorations.markings}

\tikzset{
    tangent/.style={
        decoration={
            markings,
            mark=
                at position #1
                with
                {
                    \coordinate (tangent point-\pgfkeysvalueof{/pgf/decoration/mark info/sequence number}) at (0pt,0pt);
                    \coordinate (tangent unit vector-\pgfkeysvalueof{/pgf/decoration/mark info/sequence number}) at (1,0pt);
                    \coordinate (tangent orthogonal unit vector-\pgfkeysvalueof{/pgf/decoration/mark info/sequence number}) at (0pt,1);
                }
        },
        postaction=decorate
    },
    use tangent/.style={
        shift=(tangent point-#1),
        x=(tangent unit vector-#1),
        y=(tangent orthogonal unit vector-#1)
    },
    use tangent/.default=1
}

\usepackage[disable]{todonotes}

\DeclareSymbolFont{bbold}{U}{bbold}{m}{n}
\DeclareSymbolFontAlphabet{\mathbbold}{bbold}

\newcommand*{\bmm}[1][\Lambda]{\bm{\mathsf{#1}}}
\newcommand*{\dint}[2][d]{\mathsf{d}^{#1}#2\,}
\newcommand*{\eqcom}{\, ,}
\newcommand*{\eqend}{\, .}
\newcommand*{\indmat}{\mathbbold{1}}
\newcommand*{\Lville}{\mathcal{L}}
\newcommand*{\mm}[1]{\bm{\mathrm{#1}}}
\newcommand*{\spec}[2][s]{#2^{\mathsf{#1}}} 
\newcommand*{\specind}[1][s]{\mathsf{#1}} 
\newcommand*{\vv}[1]{\bm{#1}}
\newcommand*{\Eqref}[1]{Eq.~\eqref{#1}}
\newcommand*{\Name}[1]{\textsc{#1}}

\begin{document}

\preprint{AIP/123-QED}

\title{Elasticity of disordered binary crystals}

\author{ T.~Ras, M.~Szafarczyk and M.~Fuchs}
 \email{Matthias.Fuchs@uni-konstanz.de}
\affiliation{ 
Universit\"at Konstanz, Fachbereich Physik, 78457 Konstanz, Germany
}

\date{\today}

\begin{abstract}
The properties of crystals consisting of several components can be widely 
tuned. Often solid solutions are produced, 
where substitutional or interstitional disorder determines the crystal 
thermodynamic and mechanical properties. The chemical and structural
disorder impedes the study of the elasticity of such solid solutions, since standard procedures like potential expansions cannot be applied.
We present a generalization of a density-functional based approach recently developed for one-component crystals to multi-component crystals. It yields expressions for the elastic constants valid in
solid solutions with arbitrary amounts of point defects and up to the
melting temperature. Further, both acoustic and optical phonon eigenfrequencies
can be computed in linear response from the equilibrium
particle densities and established classical density functionals.
As a proof of principle, dispersion relations are computed
for two different binary crystals: A random fcc crystal as an example for a substitutional,
and a disordered sodium chloride structure as an example of an interstitial solid solution. 
In cases where one of the components couples only weakly to the others, 
the dispersion relations develop characteristic signatures. The acoustic branches become
flat in much of the first Brillouin zone, and a crossover between acoustic and optic branches
takes place at a wavelength which can far exceed the lattice spacing.
\end{abstract}

\keywords{}
\maketitle
\section{\label{sec:Introduction} Introduction}
Solid solutions are disordered multi-component crystals, with steel  one of the most prominent examples. Solid solutions can be formed by incorporation of an atomic species into a host lattice, which can take place either by substitution, viz.~the replacement of a host particle, or by fitting into interstitial sites.  
 Such substitutional or interstitial solid solutions of two components can also be considered as disordered binary crystals, where  atoms of a species do not occur on sites of a single sublattice only, or where interstitials and vacancies arise, respectively.  
 
The local packing of different particles and entropy play an important role in the formation of multi-component crystals. The recent possibility to synthesize well characterized colloidal particles opens the possibility to study complex crystals on the microscopic  time and length scales of relevance to crystal structure selection and transport mechanisms.  The example of  binary mixtures of colloidal hard spheres has shown that already  small changes in the size ratio can lead to very different crystal structures \cite{Bartlett1992}.   
Recently, colloidal realizations of binary interstitial solid solutions have made possible direct microscopy studies of the particle motions \cite{Vermolen2009, Filion2011,Higler2013,RiosdeAnda2017}.  The interstitial sodium chloride structure  (Na$_{\chi} $Cl$_1 $) was studied. Here, one species fills a sublattice (almost) completely, while the other species is diluted and fills only a fraction $\chi<1$ of the second sublattice. While intersite defects are considered negligible in that system, a wide range of interstitial concentrations has been found both by experiment and simulation \cite{Tauber2016,Higler2017,vanderMeer2017}. 
 
Binary crystals give intuitive access to the whole richness of crystal lattice vibrations including acoustic and optical phonons.
Their dispersion relations, viz.~their polarization-dependent frequency-wavenumber relations, provide information on the elastic stability under small deformations.\\[-2.2em]
\begin{figure}[H]
 \subfloat[\label{subfig:BCC-unstrained}]{\input{BCC_unstrained.figtex}}\hfill
 \subfloat[\label{subfig:BCC-strained}]{\input{BCC_strained-isotropic.figtex}}\\
 \subfloat[\label{subfig:SC_unstrained_defects}]{\input{SC_unstrained_defects.figtex}}\hfill
 \subfloat[\label{subfig:BCC_unstrained_random}]{\input{BCC_unstrained_random.figtex}}
 \caption{Schematic samples of a binary 2-dimensional crystal in various states of order. Panel \protect\subref{subfig:BCC-unstrained} shows the classical zero-temperature ground state with the basis and the lattice constant $ a $ marked. It serves as a reference for defining particle displacements $ \spec[i_s]{\vv{u}} $, marked by green arrows in \protect\subref{subfig:BCC-strained}. Panel~\protect\subref{subfig:SC_unstrained_defects} illustrates point defects for the large species, where the definition of displacements from panel (b) breaks down.  Panel~\protect\subref{subfig:BCC_unstrained_random} shows intersite defects between large and small species.}
 \label{fig:Lattice-pictures}
 \end{figure} 
\noindent In colloidal dispersions, the vibrations are overdamped and the dispersion relations determine the wavevector dependence of density fluctuations \cite{Baumgartl2008}.  In either case, in the atomic or in the colloidal one, the acoustic branches of the dispersion relations  provide the coefficients of elasticity in the long wavelength limit. This gives access to the crystal free energies, which determine phase diagrams and thermodynamic properties.   

Theoretical developments for the determination of the thermodynamic and mechanical properties of solid solutions 
 need to overcome the difficulty of local chemical and structural  disorder.
  Classical approaches to the description of solid elasticity are based upon the $ T=0  $ equilibrium positions of particles within a unique microscopic reference configuration.\cite{born1954dynamical,Ashcroft1976}
This description is limited in accuracy or even breaks down entirely at considerable disorder.  It cannot be used for solid solutions without approximations. To see why this is, consider the illustration in Fig.~\ref{fig:Lattice-pictures}. In solid solutions, particles cannot be assigned to unique lattice sites. Displacement vectors cannot be drawn for particles without specific lattice site, and thus the displacement field cannot be defined on the atomic level. Interstitial and substitutional solid solutions are illustrated in Fig.~\ref{fig:Lattice-pictures} in panels (c) and (d), respectively.
 The single vacancy-interstitial pairs in Fig.~\ref{subfig:BCC_unstrained_random}  are called intersite defects\cite{Sanz2009, Szewczyk2015}. 
Modern approaches to the elasticity of solid solutions require mean field type of approximations in the coherent-potential approach, or choices of special quasi-random structures in supercell approaches \cite{Huang2018,Ikeda2019}. A fundamental approach at finite temperatures going beyond the harmonic approximation and including substitutional and interstitial defects is still lacking.

In order to obtain the elastic dispersion relations of solid solutions, we generalize an approach from statistical hydrodynamics, first proposed by Ernst and Szamel\cite{Szamel1993}, and recently worked out to a complete description of phonons in single-component crystals with arbitrary concentrations of point defects and temperatures up to the melting point\cite{Walz2010}. The approach is potential free and thus can handle the excluded volume interaction.
 The present work generalizes the formalism to binary crystals and includes optical phonons to the description. Two specific applications of the general formalism will be presented, one studying a random fcc crystal as example for a substitutional solid solution, and  a second one studying the disordered  interstitial sodium chloride structure realized in colloid experiments\cite{Vermolen2009}. We will obtain 
results that can be evaluated by (classical) density functional theory (DFT)\cite{Evans1979} that provides a first principles approach to the thermodynamics and structures in inhomogeneous dense systems \cite{Hansen}.

A DFT approach from the modified weighted density functional approximation has been used by \Name{Das} and \Name{Singh} on three-dimensional binary hard sphere crystals\cite{Singh2007}. The authors computed equilibrium particle densities and also vacancy concentrations of, among others, a substitutionally disordered fcc crystal. With such an equilibrium state at hand, an obvious goal is the calculation of elastic constants. An approach to that end for single-component crystals has been proposed by \Name{Jari\'{c}}~and \Name{Mohanty}\cite{Jaric1988}.
Dispersion relations within the DFT approach have been first computed for single-component crystals by \Name{Mahato}~et~al.\cite{Mahato1991}.
For binary solids, dispersion relations have been obtained for a model of alkali halides by \Name{Tosi}~and \Name{Tozzini}.\cite{Tosi1994,Tozzini1995} \todo{Cite previous alike approaches.} However, these approaches expand the thermodynamic potential around the constant liquid equilibrium densities. This leads to results for the elastic couplings in the dynamical matrix different from ours. Moreover, the coupling of the strain to the defect field is not addressed, which prevents the older methods to determine the difference between the elastic coefficients at fixed density and the  ones  at fixed defect density \cite{Fleming1976}.
 There is another  fundamental difference of previous DFT-based approaches to dispersion relations from the present one: The former require an ansatz for the relation between the microscopic particle densities $ \spec{\rho}(\vv{r}) $ entering DFT and the fields of continuum mechanics, e.g.\ for strain or displacement. Here we use the \Name{Zwanzig--Mori} formalism  (as familiar from liquids) and exact DFT relations to obtain the dynamical matrix. Approximations to the density functional -- like the \Name{Percus--Yevick} approximation well treated for liquids -- enter into that general result only on evaluation.

The paper is organized as follows: In Sect.~\ref{sec:Methods} the general formalism is developed. First, the variables whose equations of motion will be set up are determined by considering conservation laws and aspects of the spontaneous breaking of translational symmetry in crystals. Next the dynamical matrix is derived, and its symmetries and long-wavelength properties discussed.  In the last subsection \ref{subsec:Binary_HS_freezing}, technical approximations are formulated in order to evaluate the derived expressions. In Sect.~\ref{sec:Results} we obtain the structural properties and dispersion relations of two model systems, a random fcc crystal and an interstitial sodium chloride structure, in order to show the versatility of the approach. In Sect.~\ref{sec:Discussion}, we discuss the characteristics of dispersion relations
in the case where one component is only weakly localized.

\section{\label{sec:Methods} Methods}
We consider a volume $ V $ containing $ \spec{N} $, $ \specind =1,2  $ identical spherical particles of two different species at number density $ \spec{n_0}=\spec{N}/ V $. The motion of the particles with masses $ \spec{m} $ is described by a (classical) Liouville operator $ \mathcal{L} $, which includes kinetic and (internal) potential energies. 
Temperature $ T $ and the densities $ \spec{n_0} $ are chosen such that the crystalline state gives the lowest free energy and translational invariance is spontaneously broken. Long-ranged order exists and the average densities vary periodically on a common real-space lattice $ \mathbb{L} $ with corresponding reciprocal lattice $ \mathbb{G} $.
\begin{equation}
\spec{n}(\vv{r}) =
\sum_{\vv{g}\in\mathbb{G}}\spec{n}_{\vv{g}}e^{i\vv{g}\cdot\vv{r}} \;,
\end{equation}
where the order parameters $ \spec{n}_{\vv{g}} $ are the Bragg-peak amplitudes at the positions of the reciprocal-lattice  vectors~$ \vv{g} $ obtained from the instantaneous particle positions $\spec{\vv{r}}_i(t)$
\begin{equation}\label{eq:def_ng}
\spec{n}_{\vv{g}} = \frac{1}{V} \sum_{i=1}^{\spec{N}} \langle  e^{-i\vv{g}\cdot\spec{\vv{r}}_i(t)} \rangle\;. 
\end{equation}
Stationary canonical averages indicated by brackets~$ \langle\cdot\rangle $ are to be evaluated in an unstrained and equilibrated crystal.  

\subsection{\label{subsec:Microscopic_and_hydrodynamic_vars} Microscopic and hydrodynamic variables}
The lifetime of collective excitations in many-particle systems can become large at long wavelengths by a handful of fundamental physical principles. They tell us by which variables the coarse-grained dynamics is characterized. In addition to conservation laws which play that role in the hydrodynamics of fluids, in crystals spontaneously broken (translational and/or rotational) continuous symmetries yield hydrodynamic modes. They can be identified by Bogoliubov's inequality \cite{Forster1990} and lead to the approach of generalized elasticity \cite{Chaikin95}.

\subsubsection{\label{subsubsec:Conserved_Quantities} Particle, momentum and energy conservation}We define microscopic species-wise particle densities $ \spec{\rho} $ and momentum densities $ \spec{ \vv{j}}$ as
\begin{align}
\spec{\rho}(\vv{r},t) &=
\sum_{i=1}^{\spec{N}}\delta\bigl(\vv{r}-\spec{\vv{r}_i}(t)\bigr) \text{ and}\notag\\
\spec{ \vv{j}}(\vv{r},t) &=
\sum_{i=1}^{\spec{N}}\spec{\vv{p}_i}(t)\delta\bigl(\vv{r}-\spec{\vv{r}_i}(t)\bigr)\eqend\notag
\end{align}
Particles are conserved species-wise which can be expressed by continuity equations
\begin{align}\label{eq:rho_continuity}
\partial_t \spec{\rho}(\vv{k},t) &=
i\vv{k}\cdot\spec{\vv{j}}(\vv{k},t)\;,\\\notag
\text{where } 
\spec{\rho}(\vv{k},t) &= \int \frac{\dint{r}}{(2\pi)^d}e^{i\vv{k}\cdot\vv{r}}\spec{\rho}(\vv{r},t)
\end{align}
is the Fourier transform to reciprocal/momentum space.
Concerning momentum, only its \emph{sum total} over all species $ \int_V\dint{r} \vv{j}(\vv{r},t)=\textsf{const}$ is conserved whose density $ \vv{j}=
\sum_{\specind}\spec{\vv{j}} $. The conservation law can be expressed locally by a momentum continuity equation
\begin{equation}\label{eq:j_continuity}
\partial_t \vv{j}(\vv{k},t)=
i\vv{k}\cdot\vv{\sigma}(\vv{k},t)\;,
\end{equation}
where $ \vv{\sigma}(\vv{k},t) $ denotes the isothermal stress tensor. Here and henceforth, all equations of motion  are given in reciprocal space.  Equation~\eqref{eq:j_continuity} identifies $\vv{j}$ as a hydrodynamic variable and justifies
its consideration within hydrodynamics. Unlike momentum conservation, the conservation of angular momentum does not contribute to the number of hydrodynamic modes.\cite{Martin1972} Energy conservation a priori adds an additional mode. However, in the isothermal regime, energy fluctuations are determined by the momentum fluctuations alone. Hence, energy fluctuations need not be considered as an independent variable either. In colloidal systems the solvent acts as a heat bath which justifies the isothermal approximation for the colloids.

Before moving on to the spontaneously broken symmetry variables, recall that the description of optical phonons involves relative rather than collective movement of particles even in the long-wavelength limit. This motivates the inclusion of the species resolved momentum density fluctuations $\spec{\vv{j}}$ into the set of dynamical variables. Note that this is an ad-hoc generalization of conventional hydrodynamics --- just as optical phonons have no hydrodynamic counterpart.
 
\subsubsection{\label{subsubsec:Broken_Symmetries} Spontaneously broken symmetry}
The reasoning in the single species case for symmetry-restoring variables\cite{Walz2010} can be easily generalized to binary crystals. Let $ \vv{g}\neq\vv{0} $ be a reciprocal lattice vector. We make use of the following Bogoliubov inequality\cite{Wagner1966}
\begin{equation}\label{eq:Bog_Ineq}
\bigl\langle\vert \delta\spec{\rho}(\vv{g}+\vv{q}) \vert^2\bigr\rangle\geq
\frac{\bigl\vert \langle j_{\alpha}^{\ast}(\vv{k})\Lville\spec{\delta\rho}(\vv{g}+\vv{q})\rangle \bigr\vert^2}
	   {\bigl\langle\vert  \Lville j_{\alpha}(\vv{k})\vert^2  \bigr\rangle }\eqcom
\end{equation}
where ${\bf q}$ is a wavevector confined to the first Brillouin zone. Then, we
employ $ \Lville=-i\partial_t $ for the Liouville operator to the denominator of Eq.~\eqref{eq:Bog_Ineq} and insert Eq.~\eqref{eq:j_continuity}.  The isotropy approximation for the stress tensor correlations, $ \langle\sigma_{\alpha\beta}^{\ast}\sigma_{\alpha\gamma}\rangle\sim R\,\delta_{\beta\gamma} $ then yields $ \bigl\langle\vert  \Lville j_{\alpha}(\vv{k}\vert^2  \bigr\rangle\sim Rk^2$.

To continue, it will be helpful to introduce notations for the mass densities \mbox{$\spec[s]{\varrho}_0=\spec{m}\spec{n_0}$}. Concerning the numerator of Eq.~\eqref{eq:Bog_Ineq}, the particle continuity equation can be used with
\begin{equation}\label{eq:Particle_continuity}
\Lville\spec{\delta\rho}(\vv{g}+\vv{q})=\Lville\sum_{i=1}^{\spec{N}}e^{-i(\vv{g}+\vv{q})\cdot\spec{\vv{r}_{i}}}=-\frac{\vv{g}+\vv{q}}{\spec{m}}\cdot\spec{\vv{j}}(\vv{g}+\vv{q})\eqend
\end{equation}
Together with the equipartition theorem
\begin{equation}\label{eq:Equipartition}
\langle{ \spec[a]{ j}_\alpha}^\ast(\vv{k}) \spec[b]{ j}_\beta(\vv{k})\rangle=\spec[a]{\varrho}_0Vk_\textsf{B}T\delta_{\specind[ab]}\delta_{\alpha\beta}
\end{equation}
we obtain
\begin{equation}\label{Bog_Ineq_Numerator}
\langle j_{\alpha}^{\ast}(\vv{k})\Lville\spec{\delta\rho}(\vv{g}+\vv{q})\rangle=-(g+q)_\alpha Vk_\textsf{B}T\spec{n}_{\vv{g}+\vv{q}-\vv{k}}\eqend
\end{equation}
Given that $\vv{k}$ is an arbitrary wave vector, we can assume $ \vv{k}=\vv{q} $ to obain the crucial inequality
\begin{equation}\label{eq:Bog_Ineq_rewritten}
\bigl\langle\vert \delta\spec{\rho}(\vv{g}+\vv{q}) \vert^2\bigr\rangle\geq
\frac{(\vv{g}+\vv{q})^2 (k_\textsf{B}T)^2\vert \spec{n}_{\vv{g}}\vert^2V^2}{Rq^2}\varpropto q^{-2}\eqend
\end{equation}
This identifies the density fluctuations $ \delta\spec{\rho}(\vv{g}+\vv{q}) $ near the reciprocal lattice vectors as the symmetry-restoring variables. Their correlations in reciprocal space diverge for small wavevector $q$ which implies that the correlations are long range in real space, like $1/r$ in three dimensions. Hence, they have to be included in the set of slow variables in generalized elasticity \cite{Forster1990}.The special case $ \vv{g} = \vv{0} $ is included by the continuity equations~(\ref{eq:rho_continuity}).

\subsection{\label{subsec:ZM_eom} Zwanzig--Mori equations of motion}
In order to obtain the isothermal, dissipationless and linear equations of motion we apply the Zwanzig--Mori formalism \cite{Forster1990} to the set of variables identified in Sec.~\ref{subsec:Microscopic_and_hydrodynamic_vars}. 
This projection approach is known to decompose the time evolution into reversible and dissipative components, represented by the frequency matrix $ \mm{\Omega} $ respectively the memory matrix $ \mm{\Gamma} $. Besides these deterministic contributions there exists also a noise term $ \varphi_i $. The equations of motion take the following schematic form
 \begin{multline}
\partial_t \delta A_i^\ast(\vv{k},t)=-\sum_j\int\dint{k'}\bigl[ i\Omega_{ij}(\vv{k},\vv{k}')\delta A_j^\ast(\vv{k}',t) +{}\\+\int_{0}^{t}\dint[]{t}\Gamma_{ij}(\vv{k},\vv{k}',t-t')\delta A_j^\ast(\vv{k}',t') \bigr] + \varphi_i(t)\eqend
 \end{multline}
 The frequency matrix is defined as
 \begin{equation}\label{eq:FreqMat}
 \Omega_{ij}(\vv{k},\vv{k}')=\sum_k\int\dint{k''}\frac{\big\langle \delta A_i^\ast(\vv{k})\Lville\delta A_k(\vv{k}'') \big\rangle}{\big\langle \delta A_k^\ast(\vv{k}'') \delta A_j(\vv{k}')  \big\rangle}\eqend
 \end{equation}
The set of variables $ \left\lbrace A_i\right\rbrace $ employed in this work consists of the species-wise resolved momentum densities $ \spec{\vv{j}(\vv{q}) }$ and particle density flucutations $ \spec{\delta\rho}(\vv{g}+\vv{q})\eqqcolon \spec{\delta\rho_{\vv{g}}}(\vv{q}) $ for $ s=1,\ldots,B=2 $ and all reciprocal lattice vectors $ \vv{g}\in\mathbb{G} $. As we aim for the reversible dispersion relations, we only require the frequency matrix in the following. For a definition of the memory matrix $ \mm{\Gamma} $ and the noise terms $ \varphi_i $, we refer to the literature\cite{Forster1990} as their dissipative contributions are not relevant to the scope of this paper. The variables $ \delta A_i $ are to be understood as local fluctuation ensemble averages within the linear-response regime. Lars Onsager's regression hypothesis allows to describe their time evolution by the Zwanzig--Mori coupling matrices, viz \emph{equilibrium} correlation functions.\footnote{The e.o.m.\ presented are not of the type of generalized Langevin equations. The latter involve a fluctuating force term orthogonal to the subspace of slow variables. We assume this term to be negligible which certainly is true for the hydrodynamic acoustic phonon modes.} We will write $ \delta \spec{n}_{\vv{g}} \equiv\langle \spec{\delta\rho_{\vv{g}}}\rangle^{\mathsf{lr}}$ for the ensemble-averaged density fluctuations in order to distinguish them from the microscopic ones $ \spec{\delta\rho_{\vv{g}}} $. Similarly, we will also write $\delta \spec{\vv{j}}$. \footnote{While used in both a phenomenological and a linear response sense the variable prescript ``$\delta$'' indicates a local fluctuation average throughout this text.}
 
 The Liouvillean $ \Lville $ changes sign under time reversal like the momentum density fluctuations $ \spec{\delta j} $ and unlike the particle density fluctuations $ \spec{\delta n_{\vv{g}}} $. Consequently, $ \mm{\Omega} $ yields couplings only between momentum and density fluctuations.  As we are interested in the long-wavelength limit alone, the wave vector argument can be restricted to the first Brillouin zone. One can argue\cite{Ras2017} that this implies $ \vv{k}=\vv{k}'=\vv{k}''=\vv{q}\in 1^{\textsf{st}}\,\textsf{BZ} $ in Eq.~\eqref{eq:FreqMat}.  The resulting equations of motion read
 \begin{subequations}
 \begin{align}
 \label{eq:ZM_eom1}
\partial_t \spec{\delta n_{\vv{g}}}(\vv{q},t) & =\sum_{\specind',\specind''}i\left(
\frac{\bigl\langle {\spec{\delta\rho}}^\ast(\vv{g}+\vv{q})\Lville\spec[s']{j_\alpha}(\vv{q})\bigr\rangle}
{\bigl\langle{ \spec[s']{j}_\alpha}^\ast(\vv{q}) \spec[s'']{j}_\beta(\vv{q})\bigr\rangle}
\right)^\ast \spec[s'']{\delta j}(\vv{q},t)\nonumber\\
& = -i\frac{\spec{n_{\vv{g}}}}{\spec{\varrho_0}}(\vv{g}+\vv{q})\cdot\spec{\delta \vv{j}_\beta}(\vv{q},t)\; ,
\end{align}
\begin{align}
 \label{eq:ZM_eom2}
 & \partial_t \spec{\delta \vv{j}}(\vv{q},t)=\notag\\ 
 &=\sum_{\specind',\specind'',\,\vv{g}',\vv{g}}i\left(
\frac{\bigl\langle {\spec{j_\alpha}}^\ast(\vv{q})\Lville\spec[s']{\delta \rho}(\vv{g}'+\vv{q})\bigr\rangle}
{\bigl\langle{ \spec[s']{\delta \rho}}^\ast(\vv{g}'+\vv{q}) \spec[s'']{\delta \rho}(\vv{g}+\vv{q})\bigr\rangle}
\right)^\ast \spec[s'']{\delta n_{\vv{g}}}(\vv{q},t)\nonumber\\
& = -i\sum_{\specind',\,\vv{g}',\vv{g}}(\vv{g}'+\vv{q}) n_{\vv{g}'}^{\ast}{\spec[ss']{J_{\vv{g}'\vv{g}}}}^\ast(\vv{q})\spec[s']{\delta n_{\vv{g}}}(\vv{q},t)\;,
 \end{align}
 \end{subequations}
 where Equations~\eqref{eq:Equipartition} and \eqref{Bog_Ineq_Numerator} were used. The inverse density fluctuation correlation matrix $ \mm{J} $ is implicitly defined through
 \begin{equation}\label{eq:Jdef}
 Vk_\textsf{B}T\delta_{\specind[ss'']}\delta_{\vv{g}\vv{g}''}=\sum_{\specind[s'],\vv{g}'}\langle {\spec{\delta \rho}}^\ast(\vv{g}+\vv{q})\spec[s']{\delta \rho}(\vv{g}'+\vv{q})\rangle \spec[s's'']{J_{\vv{g}'\vv{g}''}}(\vv{q})\eqend
 \end{equation}
Inserting the time derivative of Eq.~\eqref{eq:ZM_eom2} into Eq.~\eqref{eq:ZM_eom1} yields an oscillator equation for the momentum density fluctuations:
\begin{align}\label{eq:wave}
 &\partial_t^2 \spec{\delta \vv{j}}(\vv{q},t)=\\\notag &=-\underbrace{\sum_{\specind',\,\vv{g}',\vv{g}}\frac{\varrho_0}{\spec[s']{\varrho_0}}(\vv{g}'+\vv{q}) {\spec{n_{\vv{g}'}}}^{\ast}{\spec[ss']{J_{\vv{g}'\vv{g}}}}^\ast(\vv{q})\spec[s']{n_{\vv{g}}}(\vv{g}+\vv{q})}_{\eqqcolon \spec[ss']{\mm{\Lambda}}(\vv{q})}\cdot \frac{\spec[s']{\delta \vv{j}}}{\varrho_0}(\vv{q},t)\eqend
\end{align}
The $ d\times d $ matrix $ \spec[ss']{\mm{\Lambda}} $ couples the momentum densities of species $ \specind $ and $ \specind[s'] $. The whole of those matrices can be written in a $ B\times d $ block matrix $ \bmm $ with

\begin{equation}\label{eq:Lambda_block}
\bmm(\vv{q})=\begin{pmatrix}
\spec[11]{\mm{\Lambda}} & \spec[12]{\mm{\Lambda}} & \cdots & \spec[1B]{\mm{\Lambda}} \\
\spec[21]{\mm{\Lambda}} & \spec[22]{\mm{\Lambda}} & \cdots &  \spec[2B]{\mm{\Lambda}} \\
\vdots & \vdots & \ddots & \vdots \\
\spec[B1]{\mm{\Lambda}} & \spec[B2]{\mm{\Lambda}} & \cdots & \spec[BB]{\mm{\Lambda}}
\end{pmatrix}(\vv{q})\eqend
\end{equation}
In Section~\ref{subsec:DynMat_properties} we discuss $ \bmm $ as given from Eq.~\eqref{eq:wave} and its eigenvalues. These will lead us to the dispersion relations and, in particular, the required $ d $ acoustic modes.

\subsection{\label{subsec:DynMat_properties} Properties of the dynamical matrix $ \bmm $}

\subsubsection{Symmetries of the inverse correlation matrix $ \spec[ss']{J_{\vv{g}\vv{g}'}}(\vv{q}) $}
The projection operator formalism and the previous identification of all relevant variables, conserved and long-range correlated ones,   has given the dynamical matrix $ \bmm $ in terms of density correlation functions. 
For statements about the properties of $ \bmm $, properties of the $ \spec[ss']{J_{\vv{g}\vv{g}'}}(\vv{q}) $ need to be known.
 To begin with, the relation
\begin{equation}
\langle {\spec{\delta \rho}}^\ast(\vv{g}+\vv{q})\spec[s']{\delta \rho}(\vv{g}'+\vv{q})\rangle=
\langle {\spec[s']{\delta \rho}}^\ast(\vv{g}'+\vv{q})\spec{\delta \rho}(\vv{g}+\vv{q})\rangle
\end{equation}
implies the self-adjointness of the inverse, i.e.\
$ \spec[ss']{J_{\vv{g}\vv{g}'}}(\vv{q})=
   {\spec[s's]{J_{\vv{g}'\vv{g}}}}^\ast(\vv{q}) $
With this it is easy to see that $ \spec[ss']{\Lambda_{\alpha\beta}}(\vv{q})=\spec{\varrho_0}/\spec[s']{\varrho_0}{\spec[s's]{\Lambda_{\beta\alpha}}}^{\ast}(\vv{q}) $.
Though not fully self-adjoint, $ \bmm $ can be easily transformed to a self-adjoint form. In rescaled momentum densities, $ \spec{\delta \vv{j}}\rightarrow \spec{\delta \bar{\vv{j}}}=\sqrt{\varrho_0/\spec{\varrho_0}} \spec{\delta \vv{j}} $, the wave equation Eq.~\eqref{eq:wave} takes the form
\begin{equation}\label{eq:sajoint_eom}
\partial_t^2 \spec{\delta \bar{\vv{j}}}(\vv{q},t)  =-\spec[ss']{\bar{\mm{\Lambda}}}(\vv{q})\cdot \frac{\spec[s']{\delta \bar{\vv{j}}}}{\varrho_0}(\vv{q},t)
\end{equation}
where $ \spec[ss']{\bar{\mm{\Lambda}}}(\vv{q}) $ is obtained from~$ \spec[ss']{\mm{\Lambda}}(\vv{q}) $ through the replacement~$ \spec[s']{\varrho_0}\rightarrow \spec[ss']{\varrho_0}\coloneqq \sqrt{\spec{\varrho_0}\spec[s']{\varrho_0}} $. The self-adjoint $ \bar{\bmm}(\vv{q})=\bar{\bmm}^\dagger (\vv{q})$ is defined as the corresponding block matrix (cf.~Eq.~\eqref{eq:Lambda_block}) whose real eigenvalues can be related to the phonon eigenfrequencies.

A means of calculating $ \spec[ss']{J_{\vv{g}\vv{g}'}}(\vv{q}) $ and with that $ \bar{\bmm} $ is obtained from the Ornstein--Zernike equation for several species \cite{Rowlinson2002}
\begin{equation}\label{eq:OZ}
\delta_{\specind[ss'']}\delta(\vv{r}-\vv{r}'')=\sum_{\specind[s']}\int\dint{r'}\langle \spec{\delta\rho}(\vv{r}) \spec[s']{\delta\rho}(\vv{r}') \rangle \spec[s's'']{C}(\vv{r}',\vv{r}'')\eqend
\end{equation}
$ \spec[ss']{C} $ is defined as the second functional derivative of the Helmholtz free energy $ \mathcal{F}= \mathcal{F}^{\textsf{id}}+\mathcal{F}^{\textsf{exc}} $ and can be split up into the ideal gas contribution and the direct correlation function $ \spec[ss']{c} $ (excess part):
\begin{subequations}\label{eq:C_decomp}
\begin{align}
\spec[ss']{C}(\vv{r},\vv{r}') &=\beta\frac{\delta^2\mathcal{F}[\spec[1]{n},\spec[2]{n},\ldots,\spec[B]{n}]}{\delta \spec[s]{n}(\vv{r})\delta \spec[s']{n}(\vv{r}')} \\ &=\delta_{\specind[ss']}\frac{\delta(\vv{r}-\vv{r}')}{\spec{n}(\vv{r})}-\spec[ss']{c}(\vv{r},\vv{r}')\eqend\label{eq:C_decomp2}
\end{align}
\end{subequations}
An important aspect of density fluctuations in crystals  was discussed by McCarley and Ashcroft \cite{McCarley1997}.  The density correlation functions show the same lattice periodicity as the equilibrium densities leading to the following representation
 \begin{equation}\label{eq:delrho_corr_Fser}
\langle \spec{\delta\rho}(\vv{r}) \spec[s']{\delta\rho}(\vv{r}') \rangle=\sum_{\vv{g}}e^{i\vv{g}\cdot \vv{R}}\spec[ss']{\delta n_{\vv{g}}}(\Delta\vv{r}')\eqend
 \end{equation}
Here we introduced center of mass and relative coordinates, $ (\vv{r}+\vv{r}')/2\eqqcolon\vv{R} $ and $ \vv{r}-\vv{r}'\eqqcolon \Delta\vv{r}' $. This can now be used to determine  the inverse density overlaps $ \spec[ss']{J_{\vv{g}\vv{g}'}}(\vv{q}) $. Note the resemblance between the left hand sides of Eqs.~\eqref{eq:OZ}~and \eqref{eq:Jdef}. They can be matched (but for the prefactor $ k_{\mathsf{B}}T $) by applying the following Fourier transform to the left hand side of \Eqref{eq:OZ}
\begin{equation}
\iint\dint{r}\dint{r''}e^{i(\vv{g}+\vv{q})\cdot\vv{r}} e^{-i(\vv{g}''+\vv{q})\cdot\vv{r}''}\delta_{\specind[ss'']}\delta(\Delta\vv{r}'')=V\delta_{\specind[ss']}\delta_{\vv{g}\vv{g}''}.
\end{equation}
Comparison of the same Fourier transform on the right hand side with the r.h.s.\ of Eq.~\eqref{eq:Jdef} yields, in conjunction with Eqs.~\eqref{eq:C_decomp}~and \eqref{eq:delrho_corr_Fser}, for the inverse of the density fluctuation correlation matrix
\begin{align}\label{eq:J_expression}
\spec[ss']{J_{\vv{g}\vv{g}'}}(\vv{q}) &= \spec[ss^{\prime\ast}]{J_{-\vv{g}-\vv{g}'}}(-\vv{q}) =\frac{k_\textsf{B}T}{V}\spec[ss']{C}(-\vv{g}-\vv{q},\vv{g}'+\vv{q})\\
												 &=\frac{k_\textsf{B}T}{V}\iint\dint{r}\dint{r'} e^{i\vv{g}\cdot\vv{r}} e^{-i\vv{g}'\cdot\vv{r}'}e^{i\vv{q}\cdot\Delta\vv{r}'}\spec[ss']{C}(\vv{r},\vv{r}')\eqend\notag
\end{align}

\subsubsection{Existence of acoustic modes}
\emph{Continuous} translational and rotational invariance of the underlying Hamiltonian of a crystallized system allow to express the gradient respectively the curl of the local density by a density integral,
\begin{subequations}\label{eq:LMBW}
\begin{align}
\vv{\nabla}\ln \spec{n}(\vv{r})
&=\sum_{\specind[s']}\int\dint{r'}\spec[ss']{c}(\vv{r},\vv{r}')\vv{\nabla}'\spec[s']{n}(\vv{r}')\eqcom\label{eq:LMBW_trans}\\
\vv{r}\times\vv{\nabla}\ln \spec{n}(\vv{r})
&=\sum_{\specind[s']}\int\dint{r'}\spec[ss']{c}(\vv{r},\vv{r}')\vv{r}'\times\vv{\nabla}'\spec[s']{n}(\vv{r}')\eqend\label{eq:LMBW_rot}
\end{align}
\end{subequations}
Going back to Lovett, Mou, Buff \cite{Lovett1976} and Wertheim \cite{Wertheim1976}, equations~\eqref{eq:LMBW_trans}~and \eqref{eq:LMBW_rot} are called the translational respectively rotational LMBW equations. More precisely, Eq.~\eqref{eq:LMBW} gives their generalization to several species systems, where we neglected a possible external potential. While that generalization has been given for the translational case\cite{Iatsevitch1997},  the straightforward derivation of the rotational LMBW-Equation~\eqref{eq:LMBW_rot} is presented in Appendix~\ref{sec:appendix_lmbw_rot}.

First, we consider again a transform in the physical coordinates:
\begin{equation}
\begin{pmatrix}
\spec[1]{\vv{j}} \\ \spec[2]{\vv{j}} 
\end{pmatrix}\rightarrow
\begin{pmatrix}
\delta\vv{j} \\ \Delta\vv{j}
\end{pmatrix}\coloneqq \frac{1}{\varrho_0}\begin{pmatrix}
									\varrho_0 & \varrho_0 \\
									\spec[2]{\varrho_0} & -\spec[1]{\varrho_0}
									\end{pmatrix}
\begin{pmatrix}
\spec[1]{\vv{j}} \\ \spec[2]{\vv{j}} 
\end{pmatrix}\eqend
\end{equation}
The transformed dynamical matrix $ \tilde{\bmm} $ takes the following form:
\begin{equation}
\tilde{\bmm}=\frac{1}{\varrho_0}\begin{pmatrix}
\sum_{\specind[s,s']}\spec[ss']{\mm{\Lambda}}\spec[s']{\varrho_0} & \spec[12]{\varrho_0}(\spec[11]{\mm{\Lambda}}-\spec[22]{\mm{\Lambda}}+\spec[21]{\mm{\Lambda}}-\spec[12]{\mm{\Lambda}})\\
\mbox{$\tilde{\mm{\Lambda}}^{\mathsf{12}}$}^\dagger &
\spec[2]{\varrho_0}\bigl(
\spec[11]{\mm{\Lambda}}+\frac{\spec[1]{\varrho_0}}{\spec[2]{\varrho_0}}\spec[22]{\mm{\Lambda}}-\spec[12]{\mm{\Lambda}}-{\spec[12]{\mm{\Lambda}}}^\dagger
\bigr)
\end{pmatrix}\eqend\label{eq:Lamb_tilde}
\end{equation}
\todo{clarify which element  $ \tilde{\bmm}^{11} $ is}
Here, one easily recognizes the top left block $\tilde{\mm{\Lambda}}^{\mathsf{11}}$ as the coupling intrinsic to the total momentum subspace, i.e.\ the coupling of the total momentum fluctuations $ \delta\vv{j} $ to themselves. The self-adjointness of $ \tilde{\bmm} $ follows again from $ \spec[ss']{\mm{\Lambda}}\equiv\spec{\varrho_0}/\spec[s']{\varrho_0}{\spec[s's]{\mm{\Lambda}}}^{\dagger} $. As usual, the eigenmodes are determined from the roots of the characteristic polynomial \mbox{$ \det[k\indmat_d-\tilde{\bmm}(\vv{q})] $}. Our approach describes acoustic modes if $ \tilde{\bmm}(\vv{q}) $ has $ d $ eigenvalue branches with leading order $ \mathcal{O}(\vv{q}^2) $ for $ \vv{q}\to\vv{0} $. Discussion of the characteristic polynomial shows that this condition is met if \emph{both} $ \spec[11]{\tilde{\mm{\Lambda}}}=\mathcal{O}(\vv{q}^2) $ and $ \spec[12]{\tilde{\mm{\Lambda}}}=\mathcal{O}(\vv{q})  $. We only show the former property from which the latter can be easily inferred.

In order to keep expressions reasonably short, we first consider only the excess terms of $ \spec[11]{\tilde{\mm{\Lambda}}} $ which stem from $ \spec[ss']{c} $ in Eq.~\eqref{eq:C_decomp2}.
\begin{multline}\label{eq:Lambda_tot_exc}
\spec[11,\,\textsf{exc}]{\tilde{\mm{\Lambda}}}(\vv{q})=-\varrho_0\sum_{s,s'}\iint \dint{r}\dint{r'} e^{-i\vv{q}\cdot\Delta\vv{r}}\spec[ss']{c}(\vv{r},\vv{r}')\times{}\\ \times(i\vv{\nabla}+\vv{q})\spec{n}(\vv{r}) (-i\vv{\nabla}'+\vv{q})\spec[s']{n}(\vv{r}')\eqend
\end{multline}
As only terms proportional to $ \vv{q}^0 $ and $ \vv{q}^1 $ need to be ruled out, the exponential in Eq.~\eqref{eq:Lambda_tot_exc} can be expanded up to linear order in $ \vv{q} $. Appendix \ref{appendixA} shows how the translational LMBW equation~\eqref{eq:LMBW_trans} can then be used to eliminate from $ \tilde{\bmm} $ all terms up to that order.

Speeds of sound and the underlying elastic constants can be more directly computed from another transform of the physical variables that block-diagonalizes the dynamical matrix. In the given long-wavelength limit $ \vv{q}\to \vv{0} $, already an approximate block diagonalization up to terms of order $ \mathcal{O}(\vv{q}^3) $ allows to drop the off-diagonal terms. It can be shown \cite[Appendix C]{Ras2017} that from the block matrix $ \tilde{\bmm} $ this leads to the following long-wavelength acoustic-modes coupling matrix  $ \spec[sound]{\mm{\Lambda}}(\vv{q}) $. With the subscript in $ \spec[ss']{\tilde{\mm{\Lambda}}_{n}} $ denoting the $ n^{\text{th}} $-order Taylor expansion coefficients,
\begin{align}
\spec[sound]{\mm{\Lambda}}(\vv{q})
&= q^2\spec[sound]{\mm{\Lambda}}_2 = q^2 \spec[11]{\tilde{\mm{\Lambda}}_2}+{}\notag\\
&+ q^2 \bigl[ (\spec[12]{\tilde{\mm{\Lambda}}_1} \cdot \spec[22^{-1}]{\tilde{\mm{\Lambda}}_0} \cdot \spec[21]{\tilde{\mm{\Lambda}}_1}) + (\ldots)^\dagger\bigr]\eqend\label{eq:Lamb_sound}
\end{align}
Note that $ \spec[12]{\tilde{\mm{\Lambda}}_1} $ and $ \spec[21]{\tilde{\mm{\Lambda}}_1} $ are of linear order in $ \vv{q} $ and thus vanish in crystals with inversion symmetry --- an argument that will be explained in more detail in Appendix~\ref{sec:Definition_Inv_Sym}.

\subsubsection{Crystal-Symmetry properties}\label{sec:Crystal-Symmetry}
As translational invariance guarantees the existence of acoustic modes, rotational invariance can be used \cite{Walz2009} to fully identify the Voigt symmetries \cite{Ashcroft1976} of the underlying elastic coefficients. The first term on the r.h.s.\ of \Eqref{eq:Lamb_sound} $ q^2 \spec[11]{\tilde{\mm{\Lambda}}_2} $ can be analyzed along the lines of the single-component case \cite[Section 3.2.2]{Walz2009}. To see this, identify
\begin{align}
q^2 \spec[11]{\tilde{\Lambda}_{2,\alpha\beta}} &=\sum_{\specind\specind[s']}\bigl( \lambda_{\alpha\beta\gamma\delta}^{\specind\specind[s']}q_\gamma q_\delta + \nu^{\specind\specind[s']} q_\alpha q_\beta+{}\bigr.\notag\\
& \bigl. {}+ q_\alpha \mu_{\beta\gamma}^{\specind\specind[s']}q_\gamma + q_\beta \mu_{\alpha\gamma}^{\specind\specind[s']}q_\gamma\bigr)
\end{align}
where
\begin{subequations}
\begin{align}
\nu^{\specind\specind[s']} &=
\frac{k_\textsf{B}T}{V}\iint\dint{r}\dint{r'}\bigl[ n(\vv{r})\delta_{\specind\specind[s']} \delta(\Delta\vv{r}')+{}\notag\\
&\bigl.{}-n^{\specind}(\vv{r})c^{\specind\specind'}(\vv{r},\vv{r}')n^{\specind'}(\vv{r}')\bigr]\eqcom\\
\mu_{\alpha\beta}^{\specind\specind[s']} &=
\frac{k_\textsf{B}T}{V}\iint\dint{r}\dint{r'}
n^{\specind}(\vv{r})c^{\specind\specind'}(\vv{r},\vv{r}')\Delta r'_\beta \nabla'_\alpha n^{\specind'}(\vv{r}')\eqcom\\\label{eq:lambda}
\lambda_{\alpha\beta\gamma\delta}^{\specind\specind[s']} &=
\frac{k_\textsf{B}T}{2V}\iint\dint{r}\dint{r'} \Delta r'_\gamma \Delta r'_\delta 
c^{\specind\specind'}(\vv{r},\vv{r}')\times {}\\\notag
&{}\times\nabla_\alpha n^{\specind}(\vv{r}) \nabla'_\beta n^{\specind'}(\vv{r}') \qquad \text{with } \Delta\vv{r}'=\vv{r}-\vv{r}'\eqend
\end{align}
\end{subequations}
 Note that both species indices are summed over in \Eqref{eq:Lamb_sound}. Making use of the \emph{rotational} LMBW equation \Eqref{eq:LMBW_rot}, it is then straightforward to generalize the one-component calculation to the case of several species. By replacing
\begin{equation}
\int\dint{r_{\nicefrac{1}{2}}} \longrightarrow \sum_{\nicefrac{\specind}{\specind[s']}}\int\dint{r/r'}
\end{equation}
and adding the species indices to $ n(\vv{r}) $, $ V(\vv{r}) $ and $ c(\vv{r},\vv{r}') $ accordingly, the index symmetries of Voigt notation are reobtained (see Appendix~\ref{appendixB} for details):
\begin{subequations}
\begin{align}
\mu_{\alpha\beta} &\coloneqq \sum_{\specind[s,s']} \spec[s,s']{\mu_{\alpha\beta}} = \mu_{\beta\alpha}\eqcom\label{eq:sym_mu}\\
\lambda_{\alpha\beta\gamma\delta} &\coloneqq \sum_{\specind[s,s']} \spec[s,s']{\lambda_{\alpha\beta\gamma\delta}} = \lambda_{\gamma\delta\alpha\beta}\label{eq:sym_lambda}\eqend
\end{align}
\end{subequations}
Concerning the second term on the r.h.s.\ of \Eqref{eq:Lamb_sound}, one can introduce a fourth-rank tensor $ \bmm[\kappa] $ such that \cite{Ras2017}
\begin{equation}
\Lambda_{\alpha\beta\gamma\delta}^{\text{sound}}q_\gamma q_\delta - \spec[11]{\tilde{\Lambda}_{\alpha\beta\gamma\delta}}q_\gamma q_\delta = \kappa_{\alpha\beta\gamma\delta} q_\gamma q_\delta\eqend\label{eq:kappa_def}
\end{equation}
and
\begin{equation}
\kappa_{\alpha\beta\gamma\delta} =
\kappa_{\gamma\beta\alpha\delta} =
\kappa_{\alpha\delta\gamma\beta} =
\kappa_{\gamma\delta\alpha\beta}\eqend\notag
\end{equation}
The symmetric summation over the indices $ \gamma  $ and $ \delta $ allows to introduce $ \tilde{\bmm[\kappa]} $ which is also symmetric w.r.t.\ interchanging the two leftmost and two rightmost indices. Thus the required Voigt symmetries also hold for $\kappa_{\alpha\beta\gamma\delta}$.  

To summarize, we have shown the correct acoustic long-wavelength behaviour of our theory which is described by a $ d $-dimensional matrix $ \spec[sound]{\mm{\Lambda}}(\vv{q}) $ obeying the Voigt symmetries required by the theory of elasticity.  Consequently, expressions for the elastic coefficients can be derived \cite{Walz2009} from direct correlation functions, which can be obtained within e.g.~density functional theory \cite{Hansen}. 
The present approach differs most importantly from the previous ones by \Name{Mahato}~et~al.\cite{Mahato1991}, \Name{Tosi}~and~\Name{Tozzini}\cite{Tosi1994,Tozzini1995} in accounting for the coupling of defect densities into the dispersion relations via the coefficients $\boldsymbol{\mu}$ and $\nu$; e.g.~the change of the stress $\boldsymbol{\sigma}$ with defect density $c$ at fixed strain $\boldsymbol{u}$ is given by\cite{Haering2015} $\left.\frac{\partial \boldsymbol{\sigma}}{\partial c}\right|_{\boldsymbol{u}}=n_0(\nu\boldsymbol{1}+\boldsymbol{\mu})$. The equilibrium density profile  remains completely general. It can contain any disorder to be described and is not restricted to an expansion around the liquid state, where the  direct correlation function would be translationally invariant,  $c({\bf r},{\bf r'})=c(|{\bf r-r'}|)$. Additionally, non-centro symmetric crystal structures can be addressed. Unlike in previous works, no ansatz for lattice displacements is required and additional couplings in non-centro symmetric structures can be considered without further ado.\todo{clarify difference to Tosi}

\subsection{\label{subsec:Binary_HS_freezing} Freezing of binary hard sphere mixtures}

While the formal theory is based on exact direct correlation functions and DFT relations, approximations are now required to evaluate the expressions for the dispersion relations. In determining the equilibrium densities $ \spec{n} $, we follow the works of Haymet and coworkers \cite{Smithline1987,McCoy1989,Rick1989,Rick1990} on the binary hard sphere freezing transition. It is based on the generalization  of the Percus--Yevick direct correlation function\cite{Percus1958,Wertheim1963} to binary systems by Lebowitz\cite{Lebowitz1964}. Based on the Helmholtz free energy equilibrium density functional $ \mathcal{F}[\spec[1]{n},\spec[2]{n}] $, the liquid-solid coexistence densities of given crystal structures are obtained as follows. The grand canonical functional $ \Delta\Omega(\spec{\mu},V,T) $ is given by
\begin{equation}
\Delta\Omega[\spec{n}]=\mathcal{F}[\spec{n}] -\mathcal{F}[\spec{n_0}] - \sum_{\specind}\spec{\mu}\int\dint{r}[\spec{n}(\vv{r})-\spec{n_0}]\eqend
\end{equation}
The reference liquid mixture state is characterized by the homogeneous particle densities $ \spec{n_0} $, the crystal by $ \spec{n} $. The functional $  \Delta\Omega $ is expanded in density deviations $ \spec{\Delta n}\equiv \spec{n}-\spec{n_0}$ about $ \spec{\Delta n}\equiv 0 $, considering all terms up to second order in $ \spec{\Delta n} $ and the central terms of third order which can be determined from
\begin{equation}
\spec[abc]{c}(\vv{k}=\vv{0},\vv{k}'=\vv{0})=\frac{\spec[ab]{\delta c}(\vv{0})}{\delta\spec[c]{n_0}}\eqend
\end{equation}
Independent of the subsequent parametrization of $ \spec{n} $, the resulting approximation for $ \Delta\Omega $ is given by
\begin{align}
&\frac{\beta}{V}\Delta\Omega=\notag\\&=\frac{\beta}{V}\sum_{\specind}\spec[id,s]{\Delta\Omega} -\frac{1}{2}\sum_{\specind[s,s']}\sum_{\vv{g}}\spec[ss']{c}(\vv{g}){\spec{\Delta n_{\vv{g}}}}^\ast\spec[s']{\Delta n_{\vv{g}}}+{}\\
&  -\frac{1}{6}\sum_{\specind[a,b,c]}\spec[abc]{c}(\vv{k}=\vv{0},\vv{k}'=\vv{0})\spec[a]{\Delta n_{\vv{0}}}\spec[b]{\Delta n_{\vv{0}}}\spec[c]{\Delta n_{\vv{0}}}\notag \eqend
\end{align}
Note that $ \spec{n_{\vv{0}}} $ with the bold subscript $ \vv{g} = \vv{0} $ stands for the particle density in the crystal which will differ from the liquid value $ \spec{n_{0}} $ at coexistence. The ideal gas contributions $ \spec[id,s]{\Delta\Omega} $ are given by the integrals
\begin{equation}\label{eq:IdGasInt}
\spec[id,s]{\Delta\Omega}=k_{\textsf{B}}T \int\dint{r}\bigl[ \spec{n}(\vv{r})\ln \frac{\spec{n}(\vv{r})}{\spec{n_0}}-\spec{\Delta n}(\vv{r}) \bigr]
\end{equation}
whose evaluation depends on the employed parameters and, more specifically, on the (de)localization of the particles at their lattice sites. We assume the widely-used Gaussian parametrization\cite{Jacobs1983}.
\begin{equation}\label{eq:parametrization}
\spec{n}(\vv{r})=\frac{\spec{\eta}}{(\sqrt{\pi} \spec{\varepsilon})^{3}}\sum_{\vv{R}\in\mathbb{L}}e^{-[(\vv{r}-\vv{R}-\spec[s]{\vv{b}})/\spec{\varepsilon}]^2}\eqend
\end{equation}
with the following parameters: $ \specind^{\textsf{th}} $ occupancy $ \spec{\eta} $ --- gives the average number of ``$ \specind $''-particles  per lattice site, Gaussian delocalization $  \spec{\varepsilon} $ --- width of the respective density peaks, Bravais lattice vectors $ \vv{R}\in\mathbb{L} $ and the $ \specind^{\textsf{th}} $ sublattice basis vectors $ \spec{\vv{b}} $. Note that generically, the density peaks are anisotropic and isotropy here is a consequence of the cubic crystal system to which all studied structures belong\cite{Lax1991}. The scalars $ \spec{\varepsilon} $ rather than matrices are then sufficient to characterize the localization of particles about their lattice sites. Evaluation of the integral~\eqref{eq:IdGasInt} can now be easily split up into different localization regimes\footnote{The characteristic length scale of the system is set by the large sphere diameter which we set to 1 arbitrary unit for ease of discussion}.
\begin{align}\label{eq:DeltaOmega_Id}
& \frac{\beta}{V}\spec[id.s]{\Delta\Omega}={}\\
& = \begin{cases}
		\spec{n_0} - \spec{n_{\vv{0}}} \bigl[\ln \spec{n_0} + 3\ln(\sqrt{\pi}\spec{\varepsilon}) + 1 + \spec{\mathcal{I}} \bigr] - \ln\spec{\eta}  & \spec{\varepsilon}\geq0.2\eqcom \\
		\spec{n_0} - \spec{n_{\vv{0}}} \bigl[\ln \spec{n_0} + 3\ln(\sqrt{\pi}\spec{\varepsilon}) + 5/2 \bigr] - \ln\spec{\eta} & \spec{\varepsilon}<0.2\eqend
	   \end{cases}\notag
\end{align}
Here,
\begin{equation}
\spec{\mathcal{I}}=\frac{\spec{\eta}}{V(\sqrt{\pi}\spec{\varepsilon})^3} \int\dint{r} \sum_{\vv{R}\in\mathbb{L}}e^{-\bigl(\frac{\vv{r}-\vv{R}}{\spec{\varepsilon}}\bigr)^2} \ln\sum_{\vv{R}'\in\mathbb{A}}e^{-\bigl(\frac{\vv{r}-(\vv{R}+\vv{R}')}{\spec{\varepsilon}}\bigr)^2}\eqend
\end{equation}
$ \mathbb{A}\subset\mathbb{L} $ denotes a set of nearest-neighbor lattice vectors whose cardinality needs to increase with increasing $ \spec{\varepsilon} $ for $ \spec{\mathcal{I}} $ to be a good approximation to the exact expression. For $ \spec{\varepsilon} $ sufficiently small, density overlap between disjoint peaks can be fully neglected, i.e.\ $ \mathbb{A}= \lbrace \vv{0} \rbrace $. The integral can then be solved analytically to give the lower line in Eq.~\eqref{eq:DeltaOmega_Id}. Details on the evaluation of $ \spec{I} $ are given in\cite[Appendix~A]{Rick1989}. The authors also present additional approximations for $ \spec{I} $ in the large $ \spec{\varepsilon} $ regime $ \spec{\varepsilon}>0.5 $. As will be seen in Sec.~\ref{subsec:Freezing_transitions}, the expressions from Eq.~\eqref{eq:DeltaOmega_Id} allow to compute most of the phase transition line.

\section{\label{sec:Results} Results}

%

With the groundwork of Section~\ref{sec:Methods}, we can now proceed to the analysis of specific disordered binary hard sphere crystal structures. The ability of treating crystals with arbitrary concentrations of point defects marks a progress for structures with considerable amounts of those defects. A structure for which this condition is met is the \emph{random fcc} structure. It is characterized in real space by a face centered cubic Bravais lattice with two identical basis vectors $ \spec[1]{\vv{b}}=\spec[2]{\vv{b}} $ that can be chosen equal to $ \vv{0} $ w.l.o.g. The statistical nature of the structure is obvious as two hard spheres cannot occupy the same lattice site simultaneously. The occupancies $ \spec{\eta} $ give the probability of finding either species $ \specind[1] $ or species $ \specind[2] $ on a given lattice site and are bounded by $ \spec[1]{\eta} + \spec[2]{\eta}\leq 1$. If they do not add up to 1 there is a certain probability of not finding a particle on a given lattice site, i.e.\ of finding a vacancy. For simplicity, we will assume complete occupancy, $ \spec[1]{\eta} + \spec[2]{\eta} =1 $, neglecting the vacancy concentration. This approximation is backed by the generally low concentration of vacancies in hard-sphere crystals.\cite{Pronk2001,Pronk2004} The Fig.~\ref{subfig:BCC_unstrained_random} gives a schematic illustration of a crystal structure with random occupancies.


\begin{figure}
 \flushleft
      \subfloat[\textsf{Random fcc coexistence densities} $ n_{\nicefrac{s}{\ell}}^{\textrm{coex}} $ \label{subfig:RandfccCoex1}]{{\input{RandomFCC_CoexistenceParams-gnuplottex-fig1.figtex}}}\\
      \subfloat[\textsf{Coexistence Gaussian widths} $ \spec{\varepsilon} $\label{subfig:RandfccCoex2}]{{\input{RandomFCC_CoexistenceParams-gnuplottex-fig2.figtex}}}
   \caption{\textsf{Coexistence parameters of the random fcc structure plotted against the HS diameter ratio~$ \varsigma $. (a) shows the particle densities of both phases scaled to the large-sphere volume on the left ordinate. The stoichiometry of the solid phase is plotted to the right ordinate. (b) shows the Gaussian localization widths. }}\label{fig:RandfccCoex}
\end{figure}

\begin{figure}
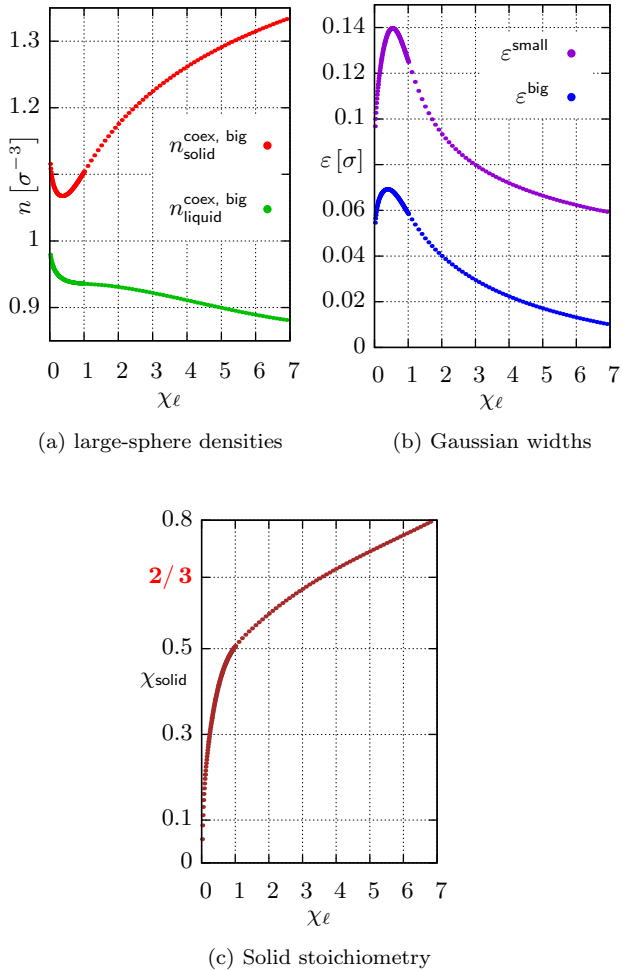

     \subfloat[large-sphere densities\label{subfig:NaxCl1_coex1}]{\input{NaxCl1_CoexistenceParams-gnuplottex-fig1.figtex}}\hspace*{.3cm} 
	 \hfil \subfloat[Gaussian widths\label{subfig:NaxCl1_coex2}]{\input{NaxCl1_CoexistenceParams-gnuplottex-fig2.figtex}}
	 \hfil \subfloat[Solid stoichiometry\label{subfig:NaxCl1_coex3}]{\input{NaxCl1_CoexistenceParams-gnuplottex-fig3.figtex}}\hfill
   \caption{Coexistence parameters of the $\textsf{Na}_{\chi}\textsf{Cl}_1$ structure as a function of the liquid stoichiometry. Diameter ratio $ \varsigma = 0.3 $.}
   \label{fig:NaxCl1_coex}
\end{figure}


As a second system, we consider an interstitial solid solution. The Na$_{\chi} $Cl$_1 $ structure has exactly the same lattice sites as a conventional sodium chloride structure. In addition to that, it shows a wide range of interstitial concentrations on its small sphere (``Na'') sublattice. The probability of finding a small hard sphere on a given Na-sublattice site rather than a vacancy\footnote{One could also take a plain large-sphere fcc structure as a reference and consider the small spheres as interstitials instead.} is given by the so-called occupancy $ \chi $ with $ 0\leq\chi\leq 1$. In this binary case with a fully occupied large-sphere sublattice, the occupancy $ \chi $ is identical to the solid stoichiometry $ \chi_{s}$. (Note that lower indices $\ell/s$ denote liquid/ solid, respectively.)  Fig.~\ref{subfig:SC_unstrained_defects} shows a spatial visualization (for better clarity, the larger spheres are randomized).
Note that unlike in the random fcc structure, we have two disjoint sublattices. The simplification here consists in a fully occupied large sphere sublattice, neglecting both vacancies and interstitials for that species.

\subsection{\label{subsec:Freezing_transitions} Freezing transitions}

We will now present liquid-solid coexistence parameters found for these two structures with the approach from Sec.~\ref{subsec:Binary_HS_freezing}. While these thermodynamic parameters were obtained already by \Name{Rick} and \Name{Haymet}\cite{Rick1989} for the random fcc crystal, the $ Na_\chi Cl_1$ structure to our knowledge has not been studied by DFT approaches. In both cases, we make use of the DFT results as input parameters for our approach to crystal dispersion relations. Any improvement in the accuracy of the DFT results will translate to more accurate dispersion relations and elastic constants.

\subsubsection{\label{subsubsec:FrTr_Rand_fcc} Random fcc structure}
Figure~\ref{subfig:RandfccCoex1} shows the liquid-solid coexistence \emph{large-sphere} densities of the random fcc structure in comparison with the previous results\cite{Rick1989}.
 For equally-sized spheres ($ \varsigma=1 $) at the least, there is physical sense to the results as we find $ 1:1 $ stoichiometry and the localization widths come to a match. As a general trend, the fluid becomes harder to crystallize as the size ratio decreases. The system compensates for this by both increasing the freezing density and decreasing the small sphere concentration in the solid. An intuitive understanding to this behavior can be obtained by thinking of the small species sites as ``partial vacancies'' which are energetically unfavorable for the crystal to maintain. Apparently, as $ \varsigma $ drops below 0.3, the small sphere Gaussians rapidly delocalize. Results for $ \varsigma<0.3 $ have not been computed as the accuracy of the Percus-Yevick input deteriorates at decreasing $\varsigma$ \cite{Hansen}.

\subsubsection{\label{subsubsec:DispRel_NaxCl1}$\textsf{Na}_{\chi}\textsf{Cl}_1$ structure}

Unlike in our first model system, the freezing transition of the $\textsf{Na}_{\chi}\textsf{Cl}_1$ structure is studied at a fixed diameter ratio $ \varsigma = 0.3 $ while the liquid stoichiometry $ \chi_{\ell} $ is varied. The results for coexisting densities, Gaussian localizations and solid stoichiometry are shown in Figures~\ref{fig:NaxCl1_coex}a, b and c, respectively. Within the large range of occupancies available from the results in Fig.~\ref{subfig:NaxCl1_coex3} we chose an approximate value of $ \chi_s\approx 2/3 $ in order to mimic the structure from experimental investigations \cite{Vermolen2009, Filion2011}. This corresponds to a stoichiometry $ \chi_{\ell} $ for the coexisting liquid of approximately $ 3.6 $. Another interesting point is situated near $ \chi_{\ell} = 0.4 $ where both the Gaussian localizations and the large-sphere solid density become extremal. This appears to be a liquid stoichiometry favorizing solidification for reasons that warrants further study.

\subsection{\label{subsec:Dispersion_Relations} Dispersion relations}

\begin{table*}[htb]
\renewcommand*{\arraystretch}{1.6}\textsf{\small{}\caption{Equilibrium crystal structure parameters employed for the dispersion relations from Fig.~\ref{fig:RandfccDispRel}}
\label{tab:HSPlotParams}}{\small \par}

\textsf{\small{}}
\begin{tabular}{|c|r@{\extracolsep{0pt}.}l|r@{\extracolsep{0pt}.}l|r@{\extracolsep{0pt}.}l|r@{\extracolsep{0pt}.}l|r@{\extracolsep{0pt}.}l|r@{\extracolsep{0pt}.}l|r@{\extracolsep{0pt}.}l|r@{\extracolsep{0pt}.}l|}
\hline 
\multirow{2}{*}{\textsf{\small{}\diagbox[width=2.8cm,innerrightsep=0.05cm,innerleftsep=0.1cm]{Structure}{Parameter}}} & \multicolumn{2}{c|}{\textsf{\small{}$\varsigma$}} & \multicolumn{2}{c|}{\textsf{\small{}$\spec[1]{m}$}} & \multicolumn{2}{c|}{\textsf{\small{}$\chi_{\ell}$}} & \multicolumn{2}{c|}{\textsf{\small{}$\spec[2]{n_{\ell}}$}} & \multicolumn{2}{c|}{\textsf{\small{}$\chi_{s}$}} & \multicolumn{2}{c|}{\textsf{\small{}$\spec[2]{n_{s}}$}} & \multicolumn{2}{c|}{\textsf{\small{}$\spec[small]{\varepsilon}$}} & \multicolumn{2}{c|}{\textsf{\small{}$\spec[big]{\varepsilon}$}}\tabularnewline
 & \multicolumn{2}{c|}{} & \multicolumn{2}{c|}{\textsf{\small{}$[m]$}} & \multicolumn{2}{c|}{} & \multicolumn{2}{c|}{\textsf{\small{}$[\sigma^{-3}]$}} & \multicolumn{2}{c|}{} & \multicolumn{2}{c|}{\textsf{\small{}$[\sigma^{-3}]$}} & \multicolumn{2}{c|}{\textsf{\small{}$[10^{-2}\sigma]$}} & \multicolumn{2}{c|}{\textsf{\small{}$[10^{-2}\sigma]$}}\tabularnewline
\hline 
\hline 
\textsf{\small{}random fcc, Fig.~\ref{subfig:RandfccDispRel1}} & \textsf{\small{}0}&\textsf{\small{}9} & \multicolumn{2}{c|}{\textsf{\small{}$\varsigma^{3}$}} & \multicolumn{2}{c|}{\textsf{\small{}1}} & \textsf{\small{}0}&\textsf{\small{}56783} & \textsf{\small{}0}&\textsf{\small{}56852} & \textsf{\small{}0}&\textsf{\small{}78768} & \textsf{\small{}7}&\textsf{\small{}8255} & \textsf{\small{}3}&\textsf{\small{}0350}\tabularnewline
\hline 
\textsf{\small{}random fcc, Fig.~\ref{subfig:RandfccDispRel2}} & \textsf{\small{}0}&\textsf{\small{}3} & \multicolumn{2}{c|}{\textsf{\small{}$\varsigma^{3}$}} & \multicolumn{2}{c|}{\textsf{\small{}1}} & \textsf{\small{}0}&\textsf{\small{}97073} & \textsf{\small{}0}&\textsf{\small{}14276} & \textsf{\small{}1}&\textsf{\small{}11065} & \textsf{\small{}39}&\textsf{\small{}371} & \textsf{\small{}2}&\textsf{\small{}1111}\tabularnewline
\hline 
\textsf{\small{}$\text{Na}_{\chi}\text{Cl}_{1}$, Fig.~\ref{fig:NaCl_DispRel}} & \textsf{\small{}0}&\textsf{\small{}3} & \multicolumn{2}{c|}{\textsf{\small{}1}} & \textsf{\small{}3}&\textsf{\small{}6} & \textsf{\small{}0}&\textsf{\small{}91533} & \textsf{\small{}0}&\textsf{\small{}66880} & \textsf{\small{}1}&\textsf{\small{}24844} & \textsf{\small{}7}&\textsf{\small{}4643} & \textsf{\small{}2}&\textsf{\small{}4853}\tabularnewline
\hline 
\end{tabular}{\small \par}
\end{table*}

The dispersion relations for the equilibrium crystal structures obtained in Sec.~\ref{subsec:Freezing_transitions} are the main result of this paper. This section focuses on presenting dispersion relations for a few model cases and explaining how to read them. We give a separate physical interpretation of their significance in Sec.~\ref{sec:Discussion}. 
 A path in the first Brillouin zone (BZ) of the fcc Bravais lattice \cite{Ashcroft1976} is chosen which contains three high symmetry directions, along which the dispersion relations and eigenvectors need to respect specific symmetry relations. 
For ease of reference, Table~\ref{tab:HSPlotParams} summarizes the equilibrium parameters of the structures for which the dispersion relations will be plotted. The particle masses of both species are proportional to the respective sphere volumes, viz particles of constant mass density are used.

\subsubsection{\label{subsubsec:DispRel_Rand_fcc} Random fcc}

\begin{figure*}
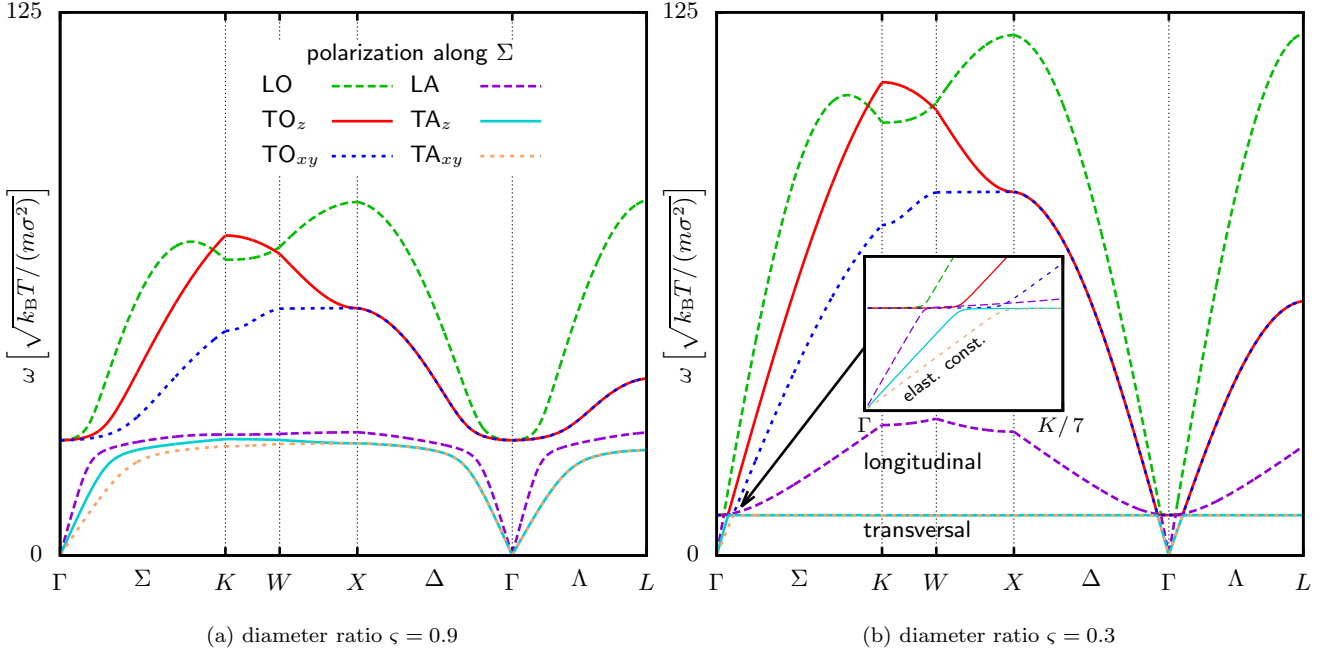

     \hfil \subfloat[diameter ratio $ \varsigma = 0.9 $ \label{subfig:RandfccDispRel1}]{\input{Disprel_RandomFCC_MassDens-DrPr-gnuplottex-fig1.figtex}}\hfil
      \subfloat[diameter ratio $ \varsigma = 0.3$ \label{subfig:RandfccDispRel2}]{\input{Disprel_RandomFCC_MassDens-DrPr-gnuplottex-fig2.figtex}}\hfill
   \caption{Dispersion relations for the random fcc structure with the equlibrium crystal density parameters given in Table~\ref{tab:HSPlotParams}. The eigenfrequencies in natural units are plotted along a path in the first BZ.}\label{fig:RandfccDispRel}
\end{figure*}

Figure~\ref{fig:RandfccDispRel} shows the dispersion relations obtained from the equilibrium parameters at size ratios $ \varsigma=0.9 $ (Fig.~\ref{subfig:RandfccDispRel1}) and $ \varsigma=0.3 $ (Fig.~\ref{subfig:RandfccDispRel2}). The plots are labeled as \textsf{L}ongitudinal respectively \textsf{T}ransversal \textsf{A}coustic and \textsf{O}ptical modes according to the eigenvector polarization along the $ \Sigma $ segment.
A couple of important features of binary dispersion relations are immediately apparent from Fig.~\ref{subfig:RandfccDispRel1}:\\
$(i)$ There are three acoustic and three optical modes showing linear respectively constant dispersion in the long-wavelength limit. Unlike for ionic crystals with their long-range interaction, the optical modes are fully degenerate at $ \vv{q}=\vv{0} $ (no LO-TO splitting).\\
$(ii)$ The transversal modes are fully degenerate along the fourfold and sixfold rotational axes of the $ X $- and $ L $-segment. Also, along the $ K $-segment that lies in the intersection of two perpendicular mirror planes, the eigenmodes occur as longitudinal and transversal ones as required by group theory.\cite{Lax1991}\\
$(iii)$ Pairwise level repulsion is visible close to the $ \Gamma $ point where acoustic and optical modes approach each other. It occurs between modes of identical polarization and is even more prominent in Fig.~\ref{subfig:RandfccDispRel2}.

While for Fig.~\ref{subfig:RandfccDispRel1}, both species are comparable in size, delocalization and concentration, Fig.~\ref{subfig:RandfccDispRel2} represents the case of spheres strongly disparate in these quantities. The small species here only has a diameter of 0.3 --- approaching the size of the tetrahedral gap in an almost close-packed large-sphere sublattice. The delocalization\todo{quadrat ?? ~$\varepsilon\approx 0.4 $ in Fig 2b! besser klammern ??(2)} $ \spec[\textsf{small}]{\varepsilon}\approx 0.4 $ clearly violates the Lindemann criterion which shows that the crystal is only stabilized by the presence of the large spheres. Still, the dispersion relations show the generic acoustic and optical long-wavelength characteristics. It is only at a critical wave vector $ q_\times $ where all of the six eigenfrequencies become approximately degenerate and two of the modes flatten to a relatively low frequency value.  If one thinks of the small species as an almost fluid background this may give a clue to these two strikingly low transversal lower modes. The onset of this effect will be referred to as the ``crossover point'' and $ q_\times $ as the ``crossover wave vector''. It also occurs on the $ \Delta $ and $ \Lambda $ segment and appears to be isotropic w.r.t.\ the direction of wave propagation.  A detailed discussion of this effect is given in Sec.~\ref{subsec:Crossover_length}.

\subsubsection{\label{subsubsec:DispRel_NaxCl2}$\textsf{Na}_{\chi}\textsf{Cl}_1$}
\begin{figure}
\input{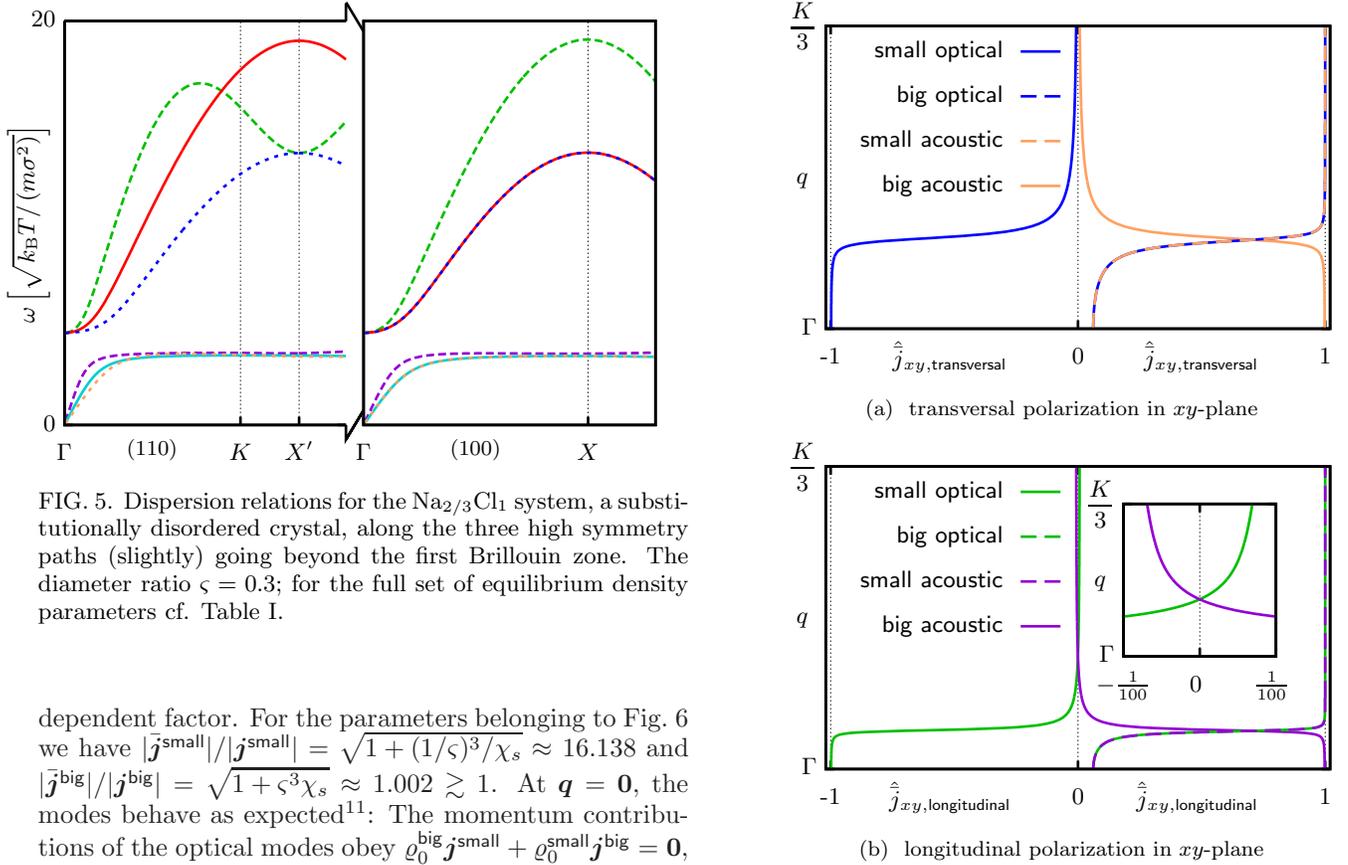}
   \caption{ Dispersion relations for the Na$_{2/3}$Cl$_1$ system, a substitutionally disordered crystal,    along the three high symmetry paths (slightly) going beyond the first Brillouin zone. The diameter ratio $ \varsigma = 0.3 $; for the full set of equilibrium density parameters cf. Table~\ref{tab:HSPlotParams}. }
   \label{fig:NaCl_DispRel}
\end{figure}

Figure~\ref{fig:NaCl_DispRel} gives the dispersion relation results for $\textsf{Na}_{\chi}\textsf{Cl}_1$ at $ \chi\approx 2/3 $. They are plotted along two of the main symmetry axes inside and slightly beyond the first Brillouin zone and show degeneracies as required by the lattice symmetries. Consider in particular the degeneracy of the transversal modes at $ X' $: The point $ X' $ marks the $ X $ point w.r.t.\ the first Brillouin zone constructed from a different, neighboring reciprocal lattice vector. \todo{clarify  following}Consequently, $ X' $ lies in turn on a four-fold rotational symmetry axis of the reciprocal lattice. Apparently, sufficient rotational symmetry enforces degeneracy of the transversal modes also beyond the first Brillouin zone.
Further note that the mirror symmetry w.r.t.\ $ X' $ respectively $ X $ proposed by the plots in Fig.~\ref{fig:NaCl_DispRel} can only be approximate. This is because of the presence of the ideal gas contribution in Eq.~\eqref{eq:C_decomp2} which depends monotonously on the magnitude of the wave vector $ \vv{q} $. The approximate validity of the mirror symmetry in the plots of Fig.~\ref{fig:NaCl_DispRel} indicates that in the given example the ideal gas term is negligible compared to the excess term. Finally, like in the random-fcc case, the lower modes do not vary within most of the range plotted. We attribute this similarity to the fact that also in this $\textsf{Na}_{\chi}\textsf{Cl}_1$ example the small-sphere species, i.e.\ the interstitials, are weakly localized compared to the large-sphere species. 

\section{\label{sec:Discussion} Discussion}

\subsection{\label{subsec:Crossover_length} Acoustic-optical crossover in random fcc}

The dispersion relations of the random fcc structure exhibit an intriguing crossover within pairs of one acoustic and one optic mode; see  Fig.~\ref{fig:RandfccDispRel}, especially  the inset of Fig.~\ref{subfig:RandfccDispRel2}.
Additional insight into this crossover can be obtained from the corresponding eigenvectors. For symmetry reasons, the directions of polarizations remain fixed across the $ \Sigma $ segment. This allows the convenient plotting of projected eigenvector components $ \hat{\bar{j}}{_{xy,\textsf{trans./long.}}} $ as shown in Fig.~\ref{fig:EigVecCrossing_xy}. Bear in mind the variable transform $ \spec{\bar{\vv{j}}}/\sqrt{\varrho_0\spec{\varrho_0}}=\spec{\vv{j}}/\spec{\varrho_0} $ that lead to the self-adjoint e.o.m.\ \eqref{eq:sajoint_eom}. Hence, $ \spec{\bar{\vv{j}}} $ equals the momentum density amplitude $ \spec{\vv{j}} $ rescaled by a constant species-dependent factor. \todo{Give factor between large and small for the plot shown} For the parameters belonging to Fig.~\ref{fig:EigVecCrossing_xy} we have $ |\spec[small]{\bar{\vv{j}}}|/ |\spec[small]{\vv{j}}| = \sqrt{1+(1/\varsigma)^3/\chi_{s}}\approx 16.138$ and $ |\spec[big]{\bar{\vv{j}}}|/ |\spec[big]{\vv{j}}| = \sqrt{1+\varsigma^3\chi_{s}}\approx 1.002\gtrsim 1$.
\begin{figure}
\smallskip

\subfloat[\label{subfig:EigVecCrossing_xyTrans} transversal polarization in $ xy $-plane]{\input{RandomFCC_EigVecTransitionGK-gnuplottex-fig1.figtex}}\hfil
\subfloat[\label{subfig:EigVecCrossing_xyLong} longitudinal polarization in $ xy $-plane]{\input{RandomFCC_EigVecTransitionGK-gnuplottex-fig2.figtex}}\hfill
\caption{Components of the normalized eigenvectors $ \hat{\bar{\vv{j}}} $ of $ \bar{\bmm} $ corresponding to 4 of the 6 modes shown in the $ \Sigma $ segment of Fig.~\ref{subfig:RandfccDispRel2}. “small/large” refers to the species considered. The color code and labelling of modes are adopted from Fig.~\ref{subfig:RandfccDispRel1}. The inset in \protect\subref{subfig:EigVecCrossing_xyLong} highlights the sign change of large acoustic and small optical component for the longitudinal mode.}
\label{fig:EigVecCrossing_xy}
\end{figure}\todo{figs side by side}
 At $ \vv{q}=\vv{0} $, the modes behave as expected \cite{Ashcroft1976}: The momentum contributions of the optical modes obey \mbox{$ \spec[big]{\varrho_0}\spec[small]{\vv{j}}+\spec[small]{\varrho_0}\spec[big]{\vv{j}}=\vv{0} $}, viz.~on average, the center of mass stays at rest. In the acoustic modes, both species contribute to the total momentum density $ \vv{j} $ according to their share in the total mass density, viz.~on average, all particles move in phase. Near the crossover wavelength, a swap in the magnitude between the small-species and large-species components of both the optical and the acoustic mode occurs.
\todo{check following}
After the crossover, the dominant \emph{momentum}\footnote{Bear in mind that momentum and displacement amplitude are related by the mass density of each species.}
contributions of small and large species to optical and acoustic mode have exchanged
their roles. The originally optical mode with largely moving small spheres is now
mainly defined by
the momentum of the large species and vice versa. This coincides
with the observation that the original optical modes in Fig.~\ref{fig:RandfccDispRel}  resemble in shape
the acoustic modes in a one-component fcc crystal\cite{Ashcroft1976}. In that
sense, the optical modes take the role of the acoustic modes in most of the
Brillouin zone, especially for weakly localized small spheres.

\begin{figure}
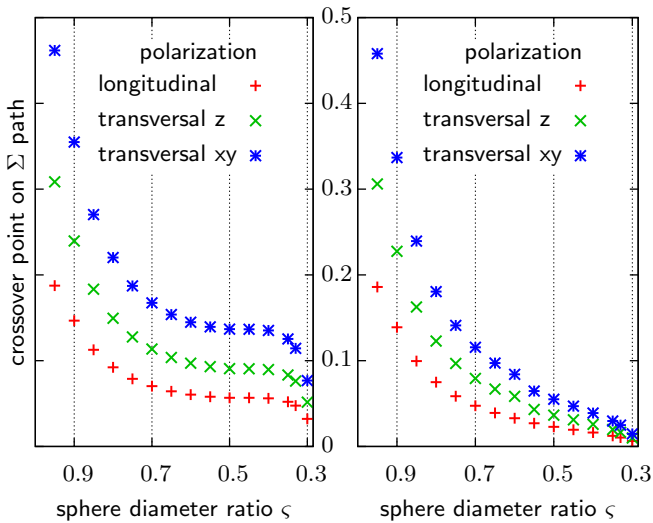

  \flushleft
      \subfloat[\textsf{Species at equal mass densities.}\label{subfig:Crossover_relative_massprop_Sigma}]{{\input{Crossover_Plots_publication_merged-gnuplottex-fig1.figtex}}}
       \subfloat[\textsf{Species at equal masses.}\label{subfig:Crossover_relative_Sigma}]{{\input{Crossover_Plots_publication_merged-gnuplottex-fig2.figtex}}}
    \caption{Relative position of the crossover wave vector $q_\times$ on the $\Sigma$ path as a function of the diameter ratio. The underlying dispersion relations were computed with particles of \protect\subref{subfig:Crossover_relative_massprop_Sigma} equal mass densities and \protect\subref{subfig:Crossover_relative_Sigma} equal masses.}\label{fig:Crossover_relative_massprop}
 \end{figure}

 In order to better understand the physical origin of the crossover, we geometrically determined its position, i.e.\ wave length $ q_\times $ in units of the $ \Sigma $ segment of the first Brillouin zone\cite{Szafarczyk2017}. 
 For each of the three acousto-optical pairs of modes in the $ \Sigma $ segment, the value of $ q_\times $ was obtained from the intersection of the linear extrapolations at $ \vv{q}=\vv{0} $, viz a degenerate constant line for the optical modes, and three different linear slopes for the acoustic modes. The result is plotted in Fig.~\ref{subfig:Crossover_relative_massprop_Sigma} as a function of the diameter ratio $ \varsigma $.\todo{Add key to these plots.}
 

 The dispersion relations from which Fig.~\ref{subfig:Crossover_relative_massprop_Sigma} was obtained are of the type of those in Fig.~\ref{fig:RandfccDispRel}, viz calculated for particles of identical mass density. Figure~\ref{subfig:Crossover_relative_Sigma} was obtained from dispersion relations with identical particle masses for the large and small species. This second plot shows a monotonous decay, viz.~increase of the crossover length with particle size asymmetry.  
As $ \varsigma $ decreases, the density profile of the small species rapidly approaches that of a homogeneous liquid background. Probing of this profile on a rather large length scale, compared to the unit cell size, still shows the characteristics of translational order --- expressed by the clear distinction into acoustic and optical modes in the inset of Fig.~\ref{subfig:RandfccDispRel2}. As one crosses $ q_\times $, moving to shorter wave lengths, the effects of translational order are lost. In particular, the lower transversal modes become very flat and we expect them to vanish completely in the limit of a homogeneous small species density. Only the lower longitudinal mode persists as expected from a homogeneous liquid. This interpretation is also backed by the apparent isotropy of the effect. Moreover, it seems plausible that the small species decouples at lower frequencies from the large species if its mass does not scale with particle volume.
In that sense, the difference between panels (a) and (b) in Fig.~\ref{fig:Crossover_relative_massprop} can be understood: In the case of particles with constant mass density, $q_x$ saturates for small $ \varsigma $ because of a balancing between the effects of decreasing localization and decreasing mass of the small species. 
 
\section{\label{sec:Conclusion} Conclusions}
\todo{check}
A theoretical  approach to obtain elastic  dispersion relations in disordered binary crystals with point disorder is presented and applied to interstitial solid solutions. It is based on the concept that acoustic phonon modes and elastic constants can be accessed from microscopically defined correlation functions starting from reciprocal space. As no microscopic reference positions are required, arbitrary amounts of point disorder can be considered. Topological defects, however, still pose a limitation to our approach. Expressions for binary crystal structures are now available which extends the scope to a broad range of common crystal structures. In the future, it will be interesting to quantify the link to the
thermodynamic description by Larch{\'e} and Cahn 
and to calculate open system elastic constants \cite{Shi2018} based on the long wavelength acoustic
limit \cite{Haering2015}. The additional finding of optical phonon modes goes beyond the theory of hydrodynamics. While these represent only some amongst a whole range of short-lived modes in the considered systems, their long-wavelength limit has sensible physical features.  The extension to crystal structures with more than two components will be straightforward.
Comparison of the theoretical results to those from colloidal model systems will provide insight into the quality of the DFT approaches employed. To that end, more up-to-date density functionals for crystals \cite{Lin2018} could be employed.

\begin{acknowledgments}
For helpful insight into their work on the $\textsf{Na}_{\chi}\textsf{Cl}_1$ structure we thank M.~Dijkstra and A.~P.~Gantapara. \mbox{M.~Oettel}, C.~Walz, F.~Miserez, J.~H\"aring, R.~Haussmann and in particular R.~Schilling supported this work by many helpful discussions. We acknowledge support from the DFG through SFB 1214 project B2.
\end{acknowledgments}

\appendix
\section{Derivation the rotational LMBW-Equation}\label{sec:appendix_lmbw_rot}
The generalization of the derivation for one-component systems\cite{Walz2010} to the case of several species is straightforward. To begin with, the coupling to external fields can be species-dependent. As a consequence, species-dependent external potentials $ \spec{\mathcal{V}}_{\textsf{ext}} $ need to be introduced for which the functional Taylor expansion reads
\begin{equation}\label{eq:Pot_variation}
\delta\spec{\mathcal{V}}_{\textsf{ext}}(\vv{r})=\sum_{\specind'}\int\dint{r'}\frac{\delta \spec{\mathcal{V}}_{\textsf{ext}}(\vv{r})}{\delta\spec[\specind']{n}(\vv{r}')}\delta\spec[\specind']{n}(\vv{r}').
\end{equation}
The invariance of the internal state of a crystal under global translations and rotations naturally extends to the case of several species. The spatial variation of a rotation $ \vv{\delta\theta} $ can be expressed by
\begin{equation}\label{eq:Rot_variation}
\vv{r}\overset{\vv{\delta\theta}}{\longrightarrow}\tilde{\vv{r}}=\vv{r}+\vv{\delta\theta}\times\vv{r}+\mathcal{O}\bigl(\vv{\delta\theta}^2\bigr).
\end{equation}
We insert the leading order of Eq.~\eqref{eq:Rot_variation} into the variation~\eqref{eq:Pot_variation},
\begin{equation}
\vv{\delta\theta}\times\vv{r}\cdot\vv{\nabla}\delta \spec{\mathcal{V}}_{\textsf{ext}}(\vv{r}) =
\sum_{\specind'}\int\dint{r'}\frac{\delta \spec{\mathcal{V}}_{\textsf{ext}}(\vv{r})}{\delta\spec[\specind']{n}(\vv{r}')}\vv{\delta\theta}\times\vv{r}\cdot\vv{\nabla}\spec[\specind']{n}(\vv{r}').
\end{equation}
In combination with the DFT relation
\begin{equation}
\beta\frac{\delta \spec{\mathcal{V}}_{\textsf{ext}}(\vv{r})}{\delta\spec[\specind']{n}(\vv{r}')}=\spec[ss']{c}(\vv{r},\vv{r}')-\delta_{\specind[ss']}\frac{\delta(\Delta\vv{r})}{\spec{n}(\vv{r})}
\end{equation}
and the cyclicity of the cross product we obtain the rotational LMBW-Equation~\eqref{eq:LMBW_rot}
\begin{align}
&\vv{r}\times\vv{\nabla}\bigl[\ln \spec{n}(\vv{r})+\beta \spec{\mathcal{V}}_{\textsf{ext}}(\vv{r})\bigr]\\\notag
&{}=\sum_{\specind[s']}\int\dint{r'}\spec[ss']{c}(\vv{r},\vv{r}')\vv{r}'\times\vv{\nabla}'\spec[s']{n}(\vv{r}').
\end{align}
This yields Equation~\eqref{eq:LMBW_rot} in the special case of vanishing external potential.
\section{Derivation of long-wavelength limit}\label{appendixA}

The proof that the dynamical matrix  $ \tilde{\bmm} $  has $d$ branches of eigenvalues which vanish like $q^2$ for long wavelength starts from Eq.~\eqref{eq:Lambda_tot_exc}. It reads, expanded up to linear order in wavevector
\begin{align}\label{eq:Lambda_tot_exc_expand}
&\frac{\spec[11,\,\textsf{exc}]{\tilde{\mm{\Lambda}}}(\vv{q})}{\varrho_0}={}\notag\\&=\sum_{\specind[s,s']}\iint \dint{r}\dint{r'} [1-i\vv{q}\cdot(\vv{r}-\vv{r}')]\spec[ss']{c}(\vv{r},\vv{r}')\times{}\notag\\ &\times \bigl[(i\vv{\nabla}+\vv{q})\spec{n}(\vv{r})i\vv{\nabla}' -i\vv{\nabla}\spec{n}(\vv{r})\vv{q}\bigr]\spec[s']{n}(\vv{r}')+\mathcal{O}(\vv{q}^2) \notag\\
&=\sum_{\specind} \int\dint{r}(i\vv{\nabla}+\vv{q})\spec{n}(\vv{r}) \times{} \\
&	\times i\sum_{\specind[s']}\int\dint{r'} \spec[ss']{c}(\vv{r},\vv{r}') \vv{\nabla}' \spec[s']{n}(\vv{r}')+{}\notag\\
& -i\sum_{\specind[s']} \int\dint{r'}\spec[s']{n}(\vv{r}')\sum_{\specind}\int\dint{r} \spec[s's]{c}(\vv{r}',\vv{r})\vv{\nabla} \spec[s]{n}(\vv{r})\vv{q}+{}\notag\\
&+i\vv{q}\cdot\sum_{\specind[s,s']}\iint \dint{r}\dint{r'}(\vv{r}-\vv{r}')\spec[ss']{c}(\vv{r},\vv{r}')\times{} \notag\\\notag
& \times \vv{\nabla}\spec{n}(\vv{r}) \vv{\nabla}'\spec[s']{n}(\vv{r}') + \mathcal{O}(\vv{q}^2)\eqend
\end{align}
In the second term of Eq.~\eqref{eq:Lambda_tot_exc_expand} we used $ \spec[ss']{c}(\vv{r},\vv{r}')=\spec[s's]{c}(\vv{r}',\vv{r}) $ which follows from the definition Eq.~\eqref{eq:C_decomp}. The first two terms now display the right hand side of Eq.~\eqref{eq:LMBW_trans}. Before we rewrite these parts, application of the same Eq.~\eqref{eq:LMBW_trans} brings the third term to vanish by
\begin{align}
&\sum_{\specind[s,s']}\iint \dint{r}\dint{r'}\vv{r}\bigl[\vv{\nabla}\spec{n}(\vv{r})\bigr]\spec[ss']{c}(\vv{r},\vv{r}')\vv{\nabla}'\spec[s']{n}(\vv{r}') ={}\notag\\
={}& \sum_{\specind} \int\dint{r}\vv{r}\bigl[\vv{\nabla}\spec{n}(\vv{r})\bigr]\frac{\vv{\nabla}\spec{n}(\vv{r})}{\spec{n}(\vv{r})}\notag \\
={}&\sum_{\specind[s,s']}\iint \dint{r}\dint{r'}\vv{r}'\bigl[\vv{\nabla}'\spec[s']{n}(\vv{r}')\bigr]\spec[s's]{c}(\vv{r}',\vv{r})\vv{\nabla}\spec{n}(\vv{r})\eqend
\end{align}
For the remaining terms, we end up with
\begin{align}
\frac{\spec[11,\,\textsf{exc}]{\tilde{\mm{\Lambda}}}(\vv{q})}{\varrho_0} &=i\sum_{\specind}\int\dint{r}\bigl[ (i\vv{\nabla}+\vv{q})\spec{n}(\vv{r})\bigr]\frac{\vv{\nabla}\spec{n}(\vv{r})}{\spec{n}(\vv{r})}+{}\notag\\
& -i\sum_{\specind}\int\dint{r}\spec{n}(\vv{r})\frac{\vv{\nabla}\spec{n}(\vv{r})}{\spec{n}(\vv{r})}\vv{q}+\mathcal{O}(\vv{q}^2) \eqend
\end{align}
Looking at Eq.~\eqref{eq:C_decomp2}, this can be easily identified as minus the ideal gas contribution in $ \spec[11]{\tilde{\mm{\Lambda}}}(\vv{q}) $ up to terms of order $ \mathcal{O}(\vv{q}^2) $. In the off-diagonal block $ \spec[12]{\tilde{\mm{\Lambda}}}(\vv{q}) = \mbox{$\tilde{\mm{\Lambda}}^{\mathsf{21}}$}^\dagger (\vv{q})$, only terms of $ \mathcal{O}(\vv{q}^0) $ can be ruled out by the same calculations as shown above. This ensures the existence of linear dispersion relations in the long-wavelength limit $ \vv{q}\to \vv{0} $. All structures considered in this paper are inversion-symmetric \todo{What do we mean by that? Note $ \bmm $ is real!} such that $ \bmm(\vv{q}) $ is even in $ \vv{q} $. In that special case, the off-diagonal blocks start out like $q^2$.  Appendix~\ref{sec:Definition_Inv_Sym} gives the specification of inversion symmetry that identifies the acoustic modes in the equations of our approach for the total momentum current.

\section{Discussion of inversion symmetry}\label{sec:Definition_Inv_Sym}
Within the context of crystals, inversion symmetry is commonly defined with reference to lattice sites as the equilibrium positions of particles. In the present approach, however, the Bravais lattice is not defined from a set of spatially periodic particle positions but only from the spatial periodicity of the equilibrium density $ n $. In that sense, it is hard to identify a privileged point within a given crystal unit cell with respect to which $ n $ can be checked for inversion symmetry. The following definition of inversion symmetry also refers only to properties of a given species-independent crystal equilibrium density $ n $ where
\begin{equation}
n(\vv{r}) =
\sum_{\vv{g}\in\mathbb{G}}n_{\vv{g}}e^{i\vv{g}\cdot\vv{r}}.
\end{equation}
We consider $ n $ as inversion symmetric in the sense of the present binary crystal approach if and only if two particle species can be distinguished with equilibrium densities $ n^{\specind[1]} $ and $ n^{\specind[2]} $ such that
$ n = n^{\specind[1]} + n^{\specind[2]} $ with
\begin{align}
n^{\specind}(\vv{r})&=\sum_{\vv{g}\in\mathbb{G}}n^{\specind}_{\vv{g}}e^{i\vv{g}\cdot\vv{r}}
\intertext{and}
\label{eq:inv_sym_prop}							& n^{\specind}_{\vv{g}} = n^{\specind}_{-\vv{g}} = {\spec{n_{\vv{g}}}}^{\ast},\; \forall\, \vv{g}\in\mathbb{G} \quad \specind = 1,2
\end{align}
where the realness of the $ n^{\specind}_{\vv{g}} $ follows from the definition~\eqref{eq:def_ng}. This means that two species equilibrium particle densities with a common reciprocal lattice and a common center of inversion can be unambiguously introduced. Note that in non-primitive crystals such as diamond this requirement can at best be approximately fulfilled.

With the property~\eqref{eq:inv_sym_prop} at hand, step by step the inversion symmetry of the dynamical matrix in the wave vector, $ \tilde{\bmm}(\vv{q}) =   \tilde{\bmm}(-\vv{q}) $,  can be inferred: Starting from the definition~\eqref{eq:delrho_corr_Fser}, immediately follows the inversion symmetry of $ \spec[ss']{C}(\vv{r},\vv{r}') $ in both its arguments. This can be used in Equation~\eqref{eq:J_expression} to obtain the realness of the density fluctuation correlation matrix, $ \spec[ss']{J_{\vv{g}\vv{g}'}} = \spec[ss^{\prime\ast}]{J_{\vv{g}\vv{g}'}} $. The inversion symmetry in $ \vv{q} $ of the dynamical matrix $ \bmm $ then follows from its definition in Equation~\eqref{eq:wave} by sign changes in the summation $ \sum_{\vv{g}\vv{g}'}\rightarrow\sum_{-\vv{g}-\vv{g}'}$ and subsequent substitution. As a consequence $ \tilde{\bmm} $ is even in $ \vv{q} $ and, as required for a long-wavelength decoupling, the off-diagonal blocks are of leading order $ \vv{q}^2 $,
\begin{equation}
\spec[21]{\tilde{\mm{\Lambda}}} = \mbox{$\tilde{\mm{\Lambda}}^{\mathsf{12}}$}^\dagger =\mathcal{O}(\vv{q}^2).
\end{equation}

\section{Derivation of Voigt symmetries}\label{appendixB}
As mentioned in the main text, in order to show the Voigt symmetries of the dynamical matrix \cite{Ashcroft1976}, we work closely along the calculations in the single-component case\cite{Walz2009}: Concerning $ \mu_{\alpha\beta} $ we combine both Equations~\eqref{eq:LMBW} as ``$ \vv{r}\times\eqref{eq:LMBW_trans}-\eqref{eq:LMBW_rot} $'' to obtain
\begin{equation}
\vv{0} = \sum_{\specind[s']}\int\dint{r'}\spec[ss']{c}(\vv{r},\vv{r}')\vv{r}'\times\vv{\nabla}'\spec[s']{n}(\vv{r}')\eqend
\end{equation}
Summation of this equation over $ \specind $ yields the symmetry Equation~\eqref{eq:sym_mu}. In order to show Eq.~\eqref{eq:sym_lambda} some tedious recombinations\cite{Walz2009} of Eq.~\eqref{eq:lambda} lead to
\begin{subequations}
\label{eq:show_sym_lambda}
\begin{align}
\label{eq:show_sym_lambda1}
&\frac{4V}{k_\textsf{B}T} \sum_{\specind[s,s']} \spec[s,s']{\lambda_{\alpha\beta\gamma\delta}} = \\\notag &= \sum_{\specind[s,s']}\iint\dint{r}\dint{r'}\spec[ss']{c}(\vv{r},\vv{r}')\Bigl( r_\delta \nabla_\gamma'\spec[s']{n}(\vv{r}') + r_\gamma \nabla_\delta'\spec[s']{n}(\vv{r}')\Bigr)\times{}\\\notag
&{}\times \Bigl(\underbrace{ r_\alpha\nabla_\beta + r_\beta\nabla_\alpha } \underline{-\,r'_\beta\nabla_\alpha- r'_\alpha\nabla_\beta}  \Bigr)\spec{n}(\vv{r})
\end{align}
and
\begin{align}
\label{eq:show_sym_lambda2}
&\frac{4V}{k_\textsf{B}T} \sum_{\specind[s,s']} \spec[s,s']{\lambda_{\gamma\delta\alpha\beta}} = \\\notag &= \sum_{\specind[s,s']}\iint\dint{r}\dint{r'}\spec[ss']{c}(\vv{r},\vv{r}')\Bigl( r_\beta \nabla_\alpha'\spec[s']{n}(\vv{r}') + r_\alpha \nabla_\beta'\spec[s']{n}(\vv{r}')\Bigr)\times{}\\\notag
&{}\times \Bigl(\underbrace{ r_\gamma\nabla_\delta + r_\delta\nabla_\gamma } \underline{-\,r'_\delta\nabla_\gamma- r'_\gamma\nabla_\delta}  \Bigr)\spec{n}(\vv{r})\eqend
\end{align}
\end{subequations}
The underlined parts in both Eqs.~\eqref{eq:show_sym_lambda} can be identified by an interchange of both the integration and the summation variables, viz $ \vv{r}\leftrightarrow\vv{r}' $ and $ \specind \leftrightarrow\specind' $. For the terms with the underbraces in Eqs.~\eqref{eq:show_sym_lambda} we employ the translational LMBW equation Eq.~\eqref{eq:LMBW_trans} in reverse direction to the primed integral and summation:
\begin{align}
&\sum_{\specind'}\int\dint\vv{r}'\spec[ss']{c}(\vv{r},\vv{r}')\Bigl( r_\beta \nabla_\alpha' + r_\alpha \nabla_\beta'\Bigr)\spec[s']{n}(\vv{r}') ={}\\\notag
{}&=r_\beta\Bigl( \frac{\nabla_\alpha \spec{n}(\vv{r})}{\spec{n}(\vv{r})} + \nabla_\alpha\spec{\mathcal{V}}_{\textsf{ext}}(\vv{r}) \Bigr)+r_\alpha\Bigl( \frac{\nabla_\beta \spec{n}(\vv{r})}{\spec{n}(\vv{r})} + \nabla_\beta\spec{\mathcal{V}}_{\textsf{ext}}(\vv{r}) \Bigr)\eqend
\end{align}
The terms with underbraces in Eqs.~\eqref{eq:show_sym_lambda} can also be identified which shows $ \lambda_{\alpha\beta\gamma\delta} = \lambda_{\gamma\delta\alpha\beta} $ and thus the symmetry of Eq.~\eqref{eq:sym_lambda}.

\section*{Disclosure of potential conflicts of interest}
The authors declare that they have no conflict of interest.
\bibliography{Draft_Tadeus_Bib}
\end{document}